\documentclass[10pt,conference]{IEEEtran}
\usepackage{subfigure}
\usepackage{multirow}
\usepackage{graphicx}
\usepackage{amsmath}
\usepackage{color}
\usepackage{epsfig}
\usepackage{amsfonts}
\usepackage{amssymb}
\usepackage{amsthm}
\usepackage[usenames,dvipsnames]{pstricks}
\usepackage{epstopdf}
\usepackage{algorithm}
\usepackage[noend]{algpseudocode}

\newtheorem{prop}{Proposition}

\newcommand{\argmin}{\mathop{\text{argmin}}}

\begin{document}

\title{{\LARGE On OTFS Modulation for High-Doppler Fading Channels}}
\author{K. R. Murali and A. Chockalingam \\
Department of ECE, Indian Institute of Science, Bangalore 560012, India}
\maketitle

\begin{abstract}
Orthogonal time frequency space (OTFS) modulation is a 2-dimensional (2D) 
modulation scheme designed in the delay-Doppler domain, unlike traditional 
modulation schemes which are designed in the time-frequency domain. Through 
a series of 2D transformations, OTFS converts a doubly-dispersive channel 
into an almost non-fading channel in the delay-Doppler domain. In this
domain, each symbol in a frame experiences an almost constant fade, thus 
achieving significant performance gains over existing modulation schemes 
such as OFDM. The sparse delay-Doppler impulse response which reflects 
the actual physical geometry of the wireless channel enables efficient 
channel estimation, especially in high-Doppler fading channels. This 
paper investigates OTFS from a signal detection and channel estimation 
perspective, and proposes a Markov chain Monte-Carlo sampling based 
detection scheme and a pseudo-random noise (PN) pilot based channel 
estimation scheme in the delay-Doppler domain.
\end{abstract}

\vspace{2mm}
{\em {\bfseries keywords:}}
{\em {\footnotesize OTFS modulation, 2D modulation, delay-Doppler domain, 
OTFS signal detection, Markov chain Monte Carlo sampling, delay-Doppler 
channel estimation. 
}} 

\section{Introduction}
\label{sec1}
\let\thefootnote\relax\footnote{This work was supported in part by the
J. C. Bose National Fellowship, Department of Science and Technology,
Government of India.}
Mobile radio channels are doubly-dispersive channels, where multipath
propagation effects cause time dispersion and Doppler shifts cause 
frequency dispersion \cite{jakes}. Multicarrier signaling schemes such 
as OFDM are often  employed to alleviate the effect of inter-symbol 
interference (ISI) caused by time dispersion \cite{ofdm1}. Doppler 
shifts result in inter-carrier interference (ICI) in OFDM which 
degrades performance \cite{ofdm2}. An approach to combat ISI and ICI 
in OFDM is pulse shaping. Pulse shaped OFDM systems use general 
time-frequency lattices and optimized pulse shapes in the time-frequency 
domain \cite{pulse1}-\cite{pulse3}. The resilience against time-frequency 
dispersions in an OFDM system depends on the time-frequency localization 
of the pulse (due to Heisenberg's uncertainty principle) and the distance 
between lattice points in the time-frequency (TF) plane \cite{pulse1}. 
Using results from sphere packing theory, \cite{pulse1} shows how to 
optimally design lattice and pulse shape for lattice-OFDM (LOFDM) 
systems in doubly-dispersive channels to jointly minimize the ISI/ICI.
A transmission scheme employing overcomplete Weyl-Heisenberg (W-H) 
frames as modulation pulses is proposed in \cite{pulse2}. In 
\cite{pulse3}, the lattice parameters and pulse shape of the 
modulation waveform are jointly optimized to adapt to the channel 
scattering function from the viewpoint of minimum symbol energy 
perturbation. While the LOFDM system in \cite{pulse1} confines the 
transmission pulses to a set of orthogonal basis functions, the pulse 
design in \cite{pulse2},\cite{pulse3} relaxed this orthogonality 
constraint in order to improve the TF concentration of the initial 
pulses, and in the process making the system more susceptible to ISI.
However, these and other such systems that employ the pulse shaping 
approach are inadequate to efficiently address the need to support 
high Doppler shifts expected in future wireless systems including 5G 
systems, where operation in high mobility scenarios (e.g., high-speed 
trains) and operation in millimeter wave (mmWave) bands are envisioned 
\cite{otfs1}. Orthogonal time frequency space (OTFS) modulation, a 
recently proposed modulation scheme \cite{otfs1}-\cite{otfs3}, has 
attractive signaling attributes that can meet the high-Doppler signaling 
need through a different approach, namely, {\em signaling in the 
delay-Doppler domain} (instead of the conventional approach of signaling 
in the time-frequency domain). In this paper, we investigate OTFS with 
emphasis on low-complexity OTFS signal detection and channel estimation 
in the delay-Doppler domain. 

OTFS waveform, having its origin from 
representation theory, is a waveform resilient to delay-Doppler shifts 
in the wireless channel \cite{otfs1}-\cite{otfs3}. The idea is to transform 
the time-varying multipath channel into a 2D channel in the delay-Doppler
domain and to carry out modulation and demodulation in this domain. Due 
to Heisenberg's uncertainty principle, a signal cannot be localized both 
in time and frequency simultaneously. But OTFS waveform is localized in 
the delay-Doppler domain, and TDMA and OFDM become limiting cases of 
OTFS when viewed in this domain. In OTFS, 2D basis functions that are 
delocalized in the time-frequency plane but are localized in the 
delay-Doppler plane are used. Information symbols are mapped onto these 
2D basis functions that span the bandwidth and time duration of the 
transmission frame. This transformation along with equalization in this 
domain makes all the symbols over a transmission frame experience the 
same channel gain, leading to good performance in high-Doppler channels. 
OTFS specializes to CDMA and OFDM if 1D basis functions (spreading codes 
and subcarriers, respectively) are used in place of 2D basis functions. 
Thus OTFS has all the advantages of TDMA, OFDM, and CDMA, and it can be 
viewed as the mother waveform of the above three. Another interesting
aspect of OTFS from an implementation view-point is that it can be 
realized by adding pre- and post-processing blocks to filtered OFDM 
systems.

\begin{figure*}
\centering
\includegraphics[width=17.5 cm, height=4.50 cm]{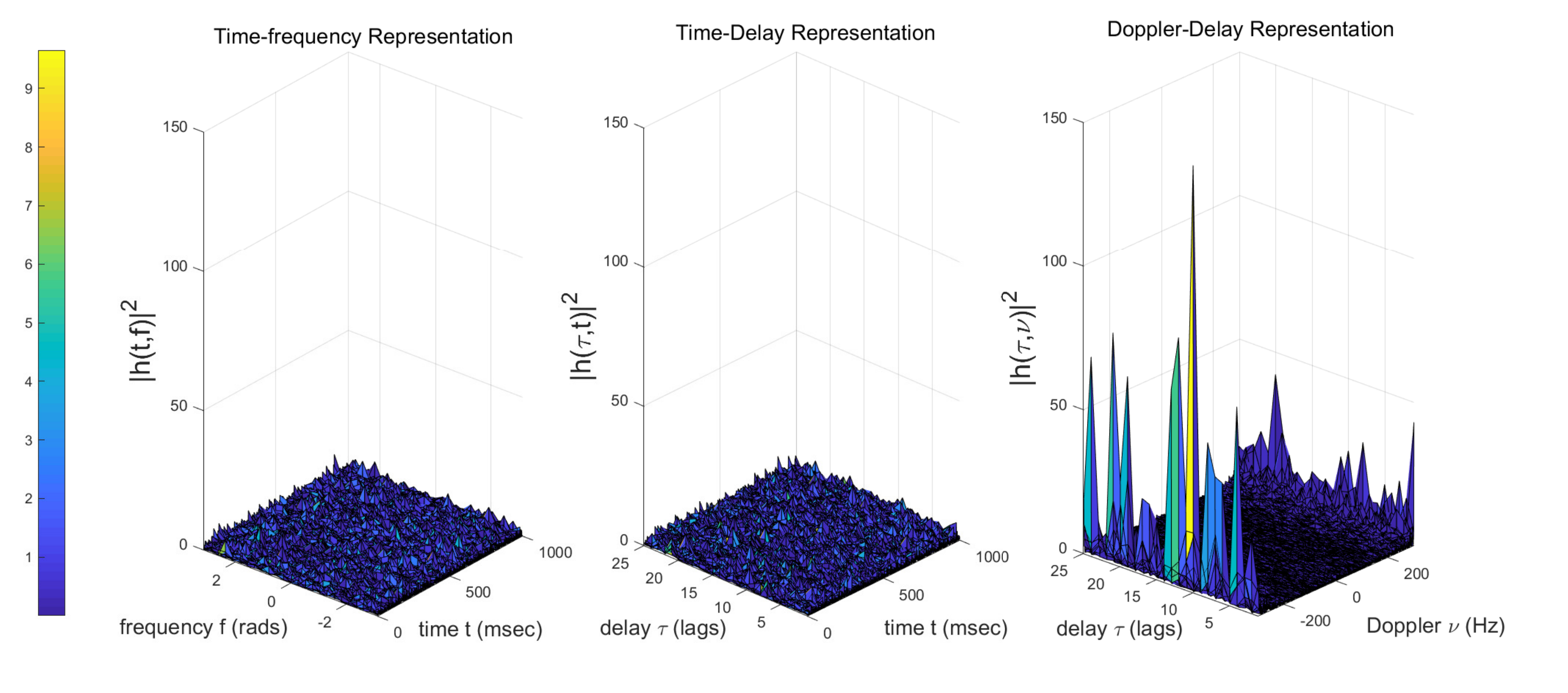}
\caption{Squared magnitude of the impulse response of a 300 Hz Jakes 
Doppler channel model with 25 uniform power delay profile taps in 
(a) time-frequency domain, (b) time-delay domain, and (c) Doppler-delay 
domain.}
\label{tdmodel}
\vspace{-2mm}
\end{figure*}

OTFS has been shown to exhibit significantly lower block error rates 
compared to OFDM over a wide range of Doppler shifts (for vehicle
speeds ranging from 30 km/h to 500 km/h in 4 GHz band), and that the 
robustness to high-Doppler channels (500 km/h vehicle speeds) is 
especially notable, as OFDM performance breaks down in such high-Doppler
scenarios \cite{otfs2}. Also, OTFS has been shown to offer bit error rate
(BER) performance advantage compared to OFDM in mmWave systems (28 GHz 
band) which encounter high frequency dispersion due to phase noise and 
high Dopplers \cite{otfs3}. An equivalent channel matrix representation
and a two-stage equalizer for OTFS are presented in \cite{otfs4}. 
Vectorized formulations of the input-output relation describing OTFS 
modulation and demodulation are presented in \cite{otfs5},\cite{otfs6}. 
A message passing based OTFS signal detection scheme based on the
vectorized formulation is presented in \cite{otfs5}. MIMO OFDM-based 
OTFS and its vectorized formulation that can enable MIMO OTFS analysis
and implementation are presented in \cite{otfs7}. Recognizing that the 
description of the OTFS waveform design framework can admit multiple 
waveforms with differences in performance depending on the delay spread 
and Doppler spread of the channel, \cite{otfs8} presents another 
modulation scheme which is robust in high-Doppler and low delay 
spread channels, termed as frequency-domain multiplexing with 
frequency-domain cyclic prefix (FDM-FDCP).

Our contribution in this paper adds to the OTFS literature that has been 
building up recently. Our contributions are two-fold. First, leveraging 
the vectorized formulations and assuming perfect knowledge of the 
equivalent channel matrix, we propose a Markov chain Monte Carlo (MCMC)
sampling based low-complexity OTFS signal detection scheme. Second, we 
relax the perfect knowledge of the equivalent channel matrix and present 
a pseudo-noise (PN) sequence pilot based channel estimation scheme in 
the delay-Doppler domain. In this context, we note that the detection 
performance of OTFS reported in the literature (e.g., 
\cite{otfs1},\cite{otfs5}) assume perfect channel knowledge.

The rest of this paper is organized as follows. The delay-Doppler channel 
representation and characteristics are presented in Sec. \ref{sec2}. The 
OTFS modulation and the linear vector system model are introduced in 
Sec. \ref{sec3}. OTFS signal detection using MCMC sampling techniques 
and the resulting bit error performance in high-Doppler scenarios are 
presented in Sec. \ref{sec4}. Channel estimation in the delay-Doppler 
domain and OTFS performance with estimated channel are presented in 
Sec. \ref{sec5}. Conclusions are presented in Sec. \ref{sec6}.

\begin{figure*}
\centering
\includegraphics[width=15.0 cm, height=4.00 cm]{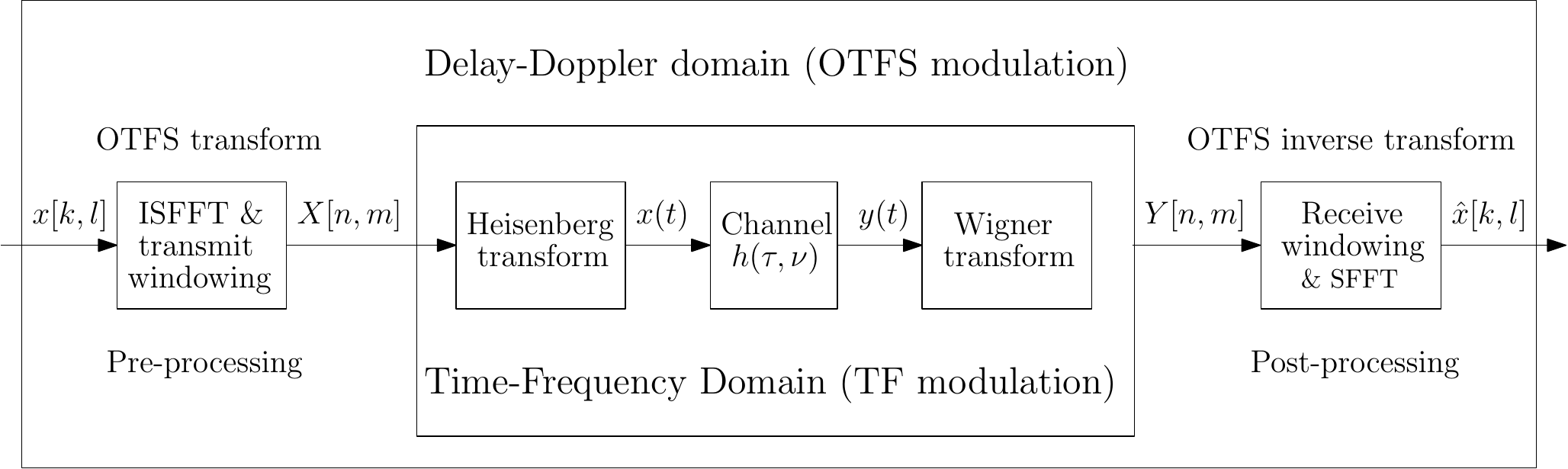}
\caption{Block diagram of OTFS modulation scheme.}
\label{fig2}
\vspace{-2mm}
\end{figure*}

\section{Wireless Channel in Delay-Doppler Domain}
\label{sec2}
When a waveform is transmitted, the wireless channel delays it in time 
(delay shift) and shifts its frequency contents (Doppler shift), and 
a delay-Doppler shifted waveform is received. A linear time-varying 
multipath channel can be represented in different ways, depending 
upon the parameters used for modeling the impulse response, namely, 
time-frequency, time-delay, Doppler-delay. Usually, time-delay
representation $h(t,\tau)$ or, equivalently, time-frequency 
representation $H(t,f)$ are used, where $t$, $\tau$, and $f$ denote 
time, delay, and frequency, respectively. These representations have 
a finite support characterized by the maximum delay and Doppler 
spreads.  The rate at which the channel coefficients vary 
($\propto 1/{\mbox{coherence time}}$) in these representations 
depends on the mobility and operating frequency. High mobility or 
high operating frequency would cause the channel to vary rapidly, 
making channel estimation and the associated time-frequency 
adaptation difficult. 

An equivalent compact way of representing the channel is to use 
delay-Doppler impulse response $h(\tau,\nu)$, where $\tau$ and $\nu$ 
denote the delay and Doppler, respectively \cite{otfs1},\cite{otfs2}. 
The taps in this domain correspond to the group of reflectors  having 
a particular delay (depends on reflectors' relative distance) and 
Doppler value (depends on reflectors' relative velocity). Thus, this 
representation reflects the actual geometry of the wireless channel 
\cite{otfs1}. Since there are only a small number of group of reflectors 
with different delay and Doppler values \cite{otfs2}, the parameters 
that need to be estimated are also fewer, and the representation in 
this domain is more compact and sparse. Also, the velocity and distance 
remain roughly the same for at least few milliseconds, and thus the 
delay-Doppler taps are time invariant for a larger observation time as 
compared to that in time-frequency representation \cite{otfs1}. This 
makes channel estimation easy in the delay-Doppler domain. 

As an illustration, we have plotted the squared magnitude of the 
impulse response of a 300 Hz (maximum Doppler) Jakes channel model 
\cite{jakes} with 25 delay taps and uniform power delay profile in 
Fig. \ref{tdmodel}; (a) in time-frequency domain, (b) in time-delay 
domain which has a Fourier transform relation with the time-frequency 
domain along the delay axis, and (c) in delay-Doppler domain which is 
related to the time-frequency domain by a transform called 2D symplectic 
Fourier transform \cite{otfs1}. As we can see from Fig. \ref{tdmodel}, 
the impulse response is not localized in the time-frequency and time-delay 
representations, whereas it is peaky (localized) in a few delay-Doppler 
bins in the delay-Doppler representation, i.e., the impulse response is 
sparse in the delay-Doppler representation. This characteristic can be 
exploited for efficient channel estimation, as we will see in Sec. 
\ref{sec5}.

In the delay-Doppler representation, the received signal $y(t)$ is the 
sum of reflected copies of the transmitted signal $x(t)$ which are 
delayed in time ($\tau$) and shifted in frequency ($\nu$) by the 
reflectors \cite{otfs1}. Thus, the coupling between an input signal 
and the channel in this domain is given by the following double integral:
\begin{equation}
\label{channel}
y(t)=\int_{\nu} \int_{\tau} h(\tau,\nu)x(t-\tau)e^{j2\pi\nu(t-\tau)} \mathrm{d} \tau \mathrm{d} \nu.
\end{equation}
While channel representation is one use of the delay-Doppler domain, 
information carrying symbols themselves can reside in this domain.
OTFS modulation is based on this \cite{otfs1},\cite{otfs2}. 

\section{OTFS Modulation}
\label{sec3}
OTFS modulation, when implemented using pre- and post-processing to 
existing multicarrier modulation schemes, can be viewed as a series of 
transformations at the transmitter and receiver. The block diagram of 
the OTFS modulation scheme is shown in Fig. \ref{fig2}, where the inner 
box is the familiar multicarrier (TF) modulation, and the outer box with 
a pre- and post-processor implements the OTFS modulation scheme in the 
delay-Doppler domain.  

The information symbols $x[k,l]$ (e.g., QAM symbols) residing in the 
delay-Doppler domain are first mapped to the familiar time-frequency 
domain symbols $X[n,m]$ through a transform called the 2D inverse 
symplectic finite Fourier transform (ISFFT) and windowing, together 
called the OTFS transform. Each $x[k,l]$ modulates a 2D basis function 
that completely spans the transmission time and bandwidth in the TF 
domain. The Heisenberg transform, which is a generalization of the OFDM 
transform, is then applied to the time-frequency transformed symbols 
$X[n,m]$ to convert to the time domain signal $x(t)$ for transmission. 
At the receiver, the received signal $y(t)$ is transformed back to a 
time-frequency domain signal $Y[n,m]$ through Wigner transform (inverse 
of the Heisenberg transform), which is a generalization of the inverse 
OFDM transform. Subsequently, $Y[n,m]$ is transformed to the delay-Doppler 
domain signal $y[k,l]$ through the symplectic finite Fourier transform
(SFFT) for demodulation. As we can see, OTFS modulation can be viewed as 
a scheme with additional pre- and post-processing to a multicarrier 
system that uses TF modulation. In the following subsections, we 
describe the signal models in TF modulation and OTFS modulation, and
present a discrete linear vector channel model of the received OTFS 
signal. 

\subsection{Time-frequency modulation and the TF lattice}
\label{sec3a}
\begin{itemize}
\item Definitions/notation: 
\begin{itemize}
\item  	A lattice in the TF plane is a sampling of the time axis at 
	an interval $T$ and the frequency axis at an interval  
	$\Delta f $ denoted by
\begin{equation*}
\hspace{-9mm}
L=\{(nT,m\Delta f), n=0,\cdots,N-1, m=0,\cdots,M-1\}.
\end{equation*}
\item A packet burst occupies $NT$ seconds in time and $M\Delta f$ Hz 
in bandwidth.
\item Information symbols $X[n,m]$, $n=0,\cdots,N-1$, $m=0,\cdots,M-1$ 
are transmitted in a given packet burst.
\item Transmit and receive pulses $\varphi_{tx}(t)$ and $\varphi_{rx}(t)$,
respectively, which are bi-orthogonal with respect to time and frequency 
translations are used for modulation and demodulation, which eliminates 
cross symbol interference.
\end{itemize}
\end{itemize}
In TF modulation, the symbols $X[n,m]$ in TF lattice are transmitted 
using the translations and modulations of the transmit pulse 
$\varphi_{tx}(t)$ as the basis functions as follows:
\begin{equation}
x(t)= \sum_{n=0}^{N-1} \sum_{m=0}^{M-1} X[n,m]\varphi_{tx}(t-nT)e^{j2\pi m \Delta f (t-nT)}.
\label{tfmod}
\end{equation}
This can be interpreted as a linear operator called the Heisenberg 
operator parametrized by $X[n,m]$ and operating on $\varphi_{tx}(t)$ . 
The transmitted signal $x(t)$ is received as $y(t)$ given by 
(\ref{channel}), which can also be viewed as another Heisenberg 
operator parametrized by $h(\tau,\nu)$ and operating on $x(t)$. Thus, 
the cascaded Heisenberg operator property can be used in the derivation 
of input-output relationship as in \cite{otfs2}.
 
In TF demodulation, the channel distorted signal $y(t)$ is matched 
filtered with the receive pulse $\varphi_{rx}(t)$ to obtain the 
sufficient statistic for detection. This is done in 2 steps. First, 
the cross ambiguity function is calculated as follows: 
\begin{equation}
\label{crossambig}
A_{\varphi_{rx},y}(\tau,\nu)=\int \varphi_{rx} ^*(t-\tau) y(t) e^{-j2 \pi \nu(t-\tau)} \mathrm{d}t.
\end{equation}
This is then sampled at an interval $\tau = nT$ and $\nu=m \Delta f$ 
to get the matched filter output, given by
\begin{equation}
\label{wigner}
Y[n,m] = A_{\varphi_{rx},y}(\tau,\nu)|_{\tau =nT,\nu =m \Delta f},
\end{equation}
where (\ref{wigner}) is called the Wigner transform.
  
Suppose the noise is additive white Gaussian, denoted by $v(t)$,
and the impulse response $h(\tau,\nu)$ has finite support bounded by 
$(\tau_{max},\nu_{max})$. If $A_{\varphi_{rx},y}(\tau,\nu) = 0 $ for 
$\tau \in (nT-\tau_{max},nT + \tau_{max}), \nu \in (m\Delta f-\nu_{max},m\Delta f + \nu_{max})$, 
then the relation between $Y[n,m]$ and $X[n,m]$ can be derived as 
\cite{otfs2}
\begin{equation}
\label{tfinpop}
Y[n,m] = H[n,m]X[n,m] + V[n,m],
\end{equation}
where $V[n,m] = A_{\varphi_{rx},v}(\tau,\nu)|_{\tau =nT,\nu =m \Delta f}$
and $H[n,m]$ is given by
\begin{equation}
H[n,m]=\int_{\tau} \int_{\nu} h(\tau,\nu) e^{j2\pi \nu nT} e^{-j2\pi (\nu + m \Delta f) \tau} \mathrm{d} \nu \mathrm{d} \tau.
\end{equation}
Clearly, each symbol $X[n,m]$ in a frame (packet burst) gets multiplied 
by a different fade $H[n,m]$ in TF modulation. However, each symbol in 
the delay-Doppler domain will be multiplied by an almost constant fade 
in the OTFS modulation presented in the following subsection. 
  
\subsection{OTFS modulation and the delay-Doppler lattice}
\label{sec3b}
The delay-Doppler signal representation is a quasi periodic representation, 
with a delay period, $\tau_r={1 \over \Delta f}$ and a Doppler period, 
$\nu_r = {1 \over T}$, such that $\tau_r \nu_r = 1$. Thus, the delay-Doppler 
lattice given below can be considered as the sampling of delay and Doppler 
axes within a rectangle of unit area ($\tau_r  \nu_r=1$) in the delay-Doppler 
domain. The delay-Doppler representation is non-unique because $\tau_r$ and 
$\nu_r$ can take any value, such that $\tau_r  \nu_r = 1$. In the limit 
$\tau_r$ tending to infinity and $\nu_r$ tending to zero, the delay-Doppler 
representation becomes the familiar temporal representation. Similarly, in 
the limit $\nu_r$ tending to infinity and $\tau_r$ tending to zero, the 
delay-Doppler representation becomes the familiar frequency domain 
representation of the signal.
\begin{itemize}
\item Definitions/notation:
\begin{itemize}
\item 	A lattice in the delay-Doppler plane is a sampling of the delay 
	axis at an interval $1  \over {M \Delta f}$ and the Doppler axis 
	at an interval  $1\over {NT}$, denoted by
{\footnotesize
\begin{equation*}
\hspace{-8mm}
L_{dD} = \left\{\left({k\over NT}, {l\over M \Delta f} \right), k=0,\cdots,N-1, l=0,\cdots,M-1 \right\}.  
\end{equation*}
}
\item 	Let $X_p[n,m]$ be the periodized  version of $X[n,m]$ with period 
	$(N,M)$. The SFFT of  $X_p[n,m]$ is then given by
\begin{equation*}
x_p[k,l] = \sum_{n=0}^{N-1} \sum_{m=0}^{M-1} X_p[n,m] e^{-j2\pi( {nk \over N} - {ml \over M} )},
\end{equation*}
and the ISFFT is $X_p[n,m] = SFFT^{-1} (x[k,l])$, given by
\begin{equation*}
X_p[n,m] = {1 \over MN }\sum_{k=0}^{N-1} \sum_{l=0}^{M-1} x[k,l] e^{j2\pi( {nk \over N}-{ml \over M})}.
\end{equation*}
\end{itemize}
\end{itemize}
 
In OTFS modulation, the information symbols in the delay-Doppler domain 
$x[k,l]$ are mapped to TF domain symbols $X[n,m]$ as 
\begin{equation}
X[n,m] = W_{tx}[n,m]SFFT^{-1}(x[k,l]),
\label{otfsmod}
\end{equation}
where $W_{tx}[n,m]$ is the transmit windowing square summable function. 
$X[n,m]$ thus obtained is in the TF domain and it is TF modulated 
as described in \ref{sec3a}, and $Y[n,m]$ is obtained by 
(\ref{crossambig}) and (\ref{wigner}).

In OTFS demodulation, a receive window $W_{rx}[n,m]$ is applied to 
$Y[n,m]$ and periodized to obtain $Y_p[n,m]$ which has the period  
$(N,M)$, as 
\begin{eqnarray}
Y_W[n,m] & = & W_{rx}[n,m]Y[n,m],  \nonumber \\
Y_p[n,m] & = & \sum_{k,l=- \infty}^{\infty}  Y_W[n-kN,m-lM].
\label{otfsdemod1}
\end{eqnarray}
The symplectic Fourier transform is then applied to $Y_p[n,m]$ to convert 
it from TF domain back to delay-Doppler domain $\hat{x}[k,l]$, as
\begin{equation}
\hat{x}[k,l]=SFFT (Y_p[n,m]).
\label{otfsdemod2}
\end{equation}
Therefore, the input-output relation in OTFS modulation can be derived 
as \cite{otfs2}
\begin{equation}
\hat{x}[k,l]={1 \over MN} \sum_{m=0}^{M-1} \sum_{n=0}^{N-1} x[n,m] h_w \left( {k-n \over NT}, {l-m \over M \Delta f} \right),
\label{otfsinpoutp}
\end{equation}
where
\begin{equation}
h_w \left({k-n \over NT}, {l-m \over M \Delta f} \right) = h_w (\nu',\tau')|_{\nu'={k-n \over NT},\tau'={l-m\over M \Delta f}},
\label{deldoppchannel}
\end{equation}
where
$h_w(\nu',\tau')$ is the circular convolution of the channel response 
with a windowing function $w(\tau,\nu)$, given by
\begin{equation}
h_w(\nu',\tau')=\int_{\nu} \int_{\tau} h(\tau,\nu)w(\nu'-\nu,\tau'-\tau) \mathrm{d} \tau \mathrm{d} \nu.
\end{equation}
The windowing function $w(\tau,\nu)$ is the symplectic discrete Fourier 
transform (SDFT) of the time frequency window 
$W[n,m]=W_{tx}[n,m]W_{rx}[n,m]$, i.e., 
\begin{equation}
w(\tau,\nu) = \sum_{m=0}^{M-1} \sum_{n=0}^{N-1} W[n,m] e^{-j2 \pi (\nu nT - \tau m \Delta f)}. 
\end{equation}

From (\ref{otfsinpoutp}), we see that each demodulated symbol $\hat{x}[k,l]$ 
for a given value of $k$ and $l$ experiences the same fade $h_w(0,0)$ on 
the transmitted symbol $x[k,l]$ and the cross symbol interference vanishes if
\begin{equation}
h_w \left({k-n \over NT}, {l-m \over M \Delta f} \right) \approx0 \: \forall n \neq k, m \neq l.
\end{equation}
This condition depends on the delay and Doppler spreads of the channel 
and the windows used in the modulation. Thus, each symbol in a given 
frame experiences an almost constant fade $h_w(0,0)$. 
  
\subsection{Vectorized formulation of the input-output relation}
\label{sec3c}
Assume that there are $P$ taps (signal propagation paths). The parameter 
$P$ is also called the sparsity of the channel. Let the path $i$ be 
associated with a delay $\tau_i$, Doppler $\nu_i$, and a fade coefficient 
$h_i$. The impulse response in the delay-Doppler domain can be written as
\begin{equation}
h(\tau,\nu) =\sum_{i=1}^{P} h_i \delta(\tau -\tau_i) \delta(\nu-\nu_i).
\label{sparsechannel}
\end{equation}
Assuming the windows used in modulation ($W_{tx}[n,m]$) and 
demodulation ($W_{rx}[n,m]$) to be rectangular, the input-output 
relation in (\ref{otfsinpoutp}) for the above channel 
can be derived as \cite{otfs5} 
\begin{eqnarray}
h_w(\tau',\nu') & \hspace{-2mm} = & \hspace{-2mm} \sum_{i=1}^{P} h_i e^{-j2 \pi \nu_i \tau_i}w(\nu' - \nu_i,\tau' - \tau_i) \nonumber \\
& \hspace{-27mm} = & \hspace{-15mm} \sum_{i=1}^{P}h_i e^{-j2 \pi \nu_i \tau_i} \sum_{c=0}^{N-1} e^{-j2 \pi (\nu'-\nu_i)cT}  \sum_{d=0}^{M-1} e^{j2 \pi (\tau'-\tau_i)d \Delta f} \nonumber \\
& \hspace{-27mm} = & \hspace{-15mm} \sum_{i=1}^{P} h_i' \mathcal{G}(\nu',\nu_i) \mathcal{F}(\tau',\tau_i), 
\end{eqnarray}
where $h_i'=h_i e^{-j2 \pi \nu_i \tau_i}$, 
$\mathcal{F}(\tau',\tau_i)=\sum_{d=0}^{M-1}e^{j2 \pi(\tau'-\tau_i)d\Delta f}$,
and $\mathcal{G}(\nu',\nu_i)= \sum_{c=0}^{N-1}e^{-j2\pi (\nu'-\nu_i)cT}$.
Define $\tau_i={ \alpha _i \over M \Delta f}$ and 
$\nu_i={(\beta_i+\gamma_i) \over NT}$, where $\alpha_i$ and $\beta_i$ are 
integers denoting the indices of the delay tap (with delay $\tau_i$) and 
Doppler tap (with Doppler value $\nu_i$), and $0 \leq \gamma_i < 1$,
where $\gamma _i $ is called the fractional Doppler which is needed 
because Doppler shifts are not exactly at the sampling points in the 
delay-Doppler plane. Now, for calculating (\ref{deldoppchannel}), 
$\mathcal{F}(\tau',\tau_i)$ is evaluated at 
$\tau'={(l-m)\over M \Delta f}$ as
\begin{eqnarray}
\mathcal{F}\left({l-m \over M \Delta f}, \tau_i \right) & = & \sum_{d=0}^{M-1} e^{j{2\pi \over M} (l-m-\alpha_i)d} \nonumber \\
& = & {e^{j2\pi(l-m-\alpha_i)}-1 \over e^{j{2 \pi \over M } (l-m-\alpha_i)}-1},
\end{eqnarray}
which evaluates to $M$ if $(l-m-\alpha_i)$ mod $M =0$ and to $0$ otherwise .
Also,
\begin{equation}
\mathcal{G}\left({k-n \over NT},\nu_i \right) =  {e^{-j2\pi(k-n-\beta_i - \gamma_i)}-1 \over e^{-j{2 \pi \over N } (k-n-\beta_i)-\gamma_i}-1}. 
\end{equation}
When ${\pi \over N} ((k-n-\beta_i) - \gamma_i)$ is small, we need to 
consider only $(2E_i+1)$ significant values of  
$\mathcal{G}\left({k-n \over NT}, \nu_i \right)$ for 
$n=(k-\beta_i+q)$ mod $N$ and $-E_i \leq q \leq E_i$, $E_i \ll N$.
 
Using the above equations, the OTFS input-output relation in  
(\ref{otfsinpoutp}) for the channel in (\ref{sparsechannel}) 
can be derived as \cite{otfs5} 
\begin{eqnarray}
\hspace{-6mm}
y[k,l] &\hspace{-2mm} = & \hspace{-2mm} \sum_{i=1}^{P} \sum_{q=-E_i}^{E_i} h_i'\left( {e^{-j2\pi(-q-\gamma_i)}-1 \over N(e^{-j{2\pi \over N}(-q-\gamma_i)}-1)} \right) \nonumber \\
& & \ \times x[((k-\beta_i+q))_N ,((l-\alpha_i))_M] + v[k,l],
\label{inpopsparse}
\end{eqnarray}
where $((.))_N$ denotes modulo $N$ operation and 
$v[k,l] \sim \mathcal{CN}(0,\sigma^2) $ is the additive Gaussian noise 
(AWGN). The fractional Doppler $\gamma_i$ affects the neighboring 
Doppler taps ($-E_i$ to $E_i$) in (\ref{inpopsparse}). This interference 
is called the inter-Doppler interference (IDI). When the Doppler taps are 
assumed to be integer multiples $(\gamma_i=0)$, (\ref{inpopsparse}) 
simplifies to
\begin{equation}
y[k,l] = \sum_{i=1}^{P} h_i' x[((k-\beta_i))_N,((l-\alpha_i))_M)] + v[k,l]. 
\label{inpopnofracdopp}
\end{equation}
We use (\ref{inpopsparse}) for OTFS signal detection in Sec. \ref{sec4} 
and (\ref{inpopnofracdopp}) for channel estimation in the delay-Doppler 
domain in Sec. \ref{sec5}. The equations (\ref{inpopsparse}) and 
(\ref{inpopnofracdopp}) can be represented in vectorized form as 
\cite{otfs5}
\begin{equation}
\mathbf{y} = \mathbf{Hx} + \mathbf{v}, 
\label{vecform}
\end{equation}
where $\mathbf{x}, \mathbf{y}, \mathbf{v} \in \mathbb{C} ^{NM \times 1}$, 
$\mathbf{H} \in \mathbb{C}^{NM\times NM}$, the $(k + Nl)$th element of 
$\mathbf{x}$,  $x_{k+Nl}=x[k,l]$, $k=0,\cdots,N-1, l=0,\cdots,M-1$, and 
the same relation holds for $\mathbf{y}$ and $\mathbf{z}$ as well. In 
this representation, there are only $P(2E_i + 1)$ non-zero elements in 
each row and column of $\mathbf{H}$ due to modulo operations. 
 
\section{OTFS Signal Detection}
\label{sec4}
In this section, we present OTFS signal detection algorithms using 
MCMC sampling based techniques.

\subsection{OTFS signal detection using MCMC sampling} 
MCMC techniques are computational techniques commonly used for calculating 
complex integrals by expressing the integral as an expectation of some 
probability distribution and then estimating this expectation from the 
generated samples of that distribution. In this technique, a new sample 
value is generated randomly from the most recent sample. The transition 
probabilities between samples are only a function of the previous sample, 
and hence the name MCMC. MCMC techniques have been used for signal detection
in multiuser and MIMO systems \cite{mcmc1}-\cite{mcmc4}. Gibbs sampling 
is a well known MCMC technique. Here, we present low-complexity Gibbs 
sampling based algorithms for OTFS signal detection using the vectorized 
formulation of the OTFS signal in (\ref{vecform}). 

\subsubsection{Gibbs sampling based OTFS detection}
\label{sec4a}
Let ${\mathbb A}$ denote the modulation alphabet used (e.g., BPSK, QAM). 
Assuming that the input symbols are equally likely, the maximum likelihood 
(ML) decision rule for the signal model in (\ref{vecform}) is given by
\begin{equation}
\label{MLGS}
\hat{\mathbf{x}}_{{\tiny \mbox{ML}}}=\argmin_{\mathbf{x} \in \mathbb{A}^{NM}} \| \mathbf{y}-\mathbf{H} \mathbf{x}\|^2,
\end{equation}
which has an exponential complexity in $NM$. However, approximate solutions
to (\ref{MLGS}) can be obtained efficiently using Gibbs sampling based MCMC 
techniques. The joint probability distribution of interest for detection is
\begin{equation}
\label{probdist}
p(x_1,x_2,\cdots,x_{NM}| \mathbf{y} , \mathbf{H}) \propto \text{exp} \left( - {|| \mathbf{y} - \mathbf{H} \mathbf{x} ||^2 \over \sigma^2 } \right).
\end{equation}
Let $t$ denote the iteration index and $k$ denote the coordinate index of 
$\mathbf{x}$. A random initial vector, denoted by $\mathbf{x}^{(t=0)} $, 
is chosen for the algorithm to begin. In every iteration, all the $NM$ 
coordinates are updated and the update in the $(t+1)$th iteration is 
obtained by sampling from the following distributions:
\begin{align*}
\label{CGS}
x_1 ^ {(t+1)} & \sim    p \big( x_1 | x_2^{(t)}, x_3^{(t)}, \cdots , x_{NM}^{(t)},\mathbf{y} , \mathbf{H}  \big ) \\
x_2 ^ {(t+1)}  & \sim    p \big( x_2 | x_1^{(t+1)}, x_3^{(t)}, \cdots , x_{NM}^{(t)} , \mathbf{y}, \mathbf{H} \big)\\
   x_3 ^ {(t+1)}  & \sim    p \big( x_3 | x_1^{(t+1)}, x_2^{(t+1)}, x_4^{(t)}, \cdots , x_{NM}^{(t)} , \mathbf{y} , \mathbf{H} \big)\\
\vdots\\
x_{NM}^{(t+1)} &  \sim   p \big( x_{NM} | x_1^{(t+1)}, x_2^{(t+1)}, \cdots , x_{NM-1}^{(t+1)},\mathbf{y} , \mathbf{H}  \big).
\end{align*}
The solution vector thus obtained in the $t$th iteration is passed to 
the $(t+1)$th iteration for the next set of coordinate updates. After a 
certain number of iterations called the burn-in period, the distribution 
tends to converge to the stationary distribution (\ref{probdist}), which 
is the distribution to simulate for drawing samples. The detected symbol 
vector in a given iteration is chosen to be that vector which has the 
least ML cost 
$f_{{\tiny \mbox{ML}}}(\mathbf{x})=\|\mathbf{y}-\mathbf{H} \mathbf{x}\|^2$ 
in all the iterations. However, Gibbs sampling algorithm suffers from a 
problem called stalling, which degrades the BER performance at high SNRs
\cite{mcmc1}. To alleviate the stalling problem and to reduce the number 
of iterations, a slightly different distribution can be used for sampling 
as described in the following subsection.

\subsubsection{Gibbs sampling with temperature parameter $\alpha$}
\label{sec_Gibbs_temp}
The expected number of iterations for finding the solution can be reduced
by using a temperature parameter $\alpha$ in the distribution as follows:
\begin{equation}
\label{probdist_alpha}
p(x_1,x_2,\cdots,x_{NM}| \mathbf{y} , \mathbf{H})  \propto \text{exp} \left(-{\|\mathbf{y}-\mathbf{H} \mathbf{x} \|^2 \over \alpha^2 \sigma^2 } \right).
\end{equation}
However, the value of $\alpha$ to be used is dependent on the operating 
SNR \cite{mcmc3}. Another possible modification of the Gibbs sampling 
to alleviate the stalling problem is to combine distributions with 
$\alpha=1$ (conventional) and $\alpha= \infty$ (uniform) using a 
randomized update rule as given in the next subsection.
 
\subsubsection{Randomized Gibbs sampling based detection}
\label{sec4b}
\begin{algorithm}[t]
\caption{Randomized Gibbs sampling based detection algorithm}
\begin{algorithmic}[1]
\State \textbf{Inputs}: $\mathbf{y}$, ${\bf H}$,
$\mathbf{x}^{(t=0)} \in \mathbb{A}^{NM}$: random initial vector, 
$ N_{iter}$: maximum no. of iterations
\State \textbf{Initialize}: $t=0$; $\mathbf{z}=\mathbf{x}^{(t=0)}$; 
index set $S=\{ 1,2,\cdots,NM\}$
\State   \textbf{Calculate} $\beta = f_{{\tiny \mbox{ML}}}(\mathbf{x}^{(t=0)})$: initial ML cost
\vspace*{2 mm}
\State  \textbf{while} ( $t < N_{iter} $) \textbf{do}
\State \hspace*{0.2 cm} \textbf{for} $k = 1$ to $NM$ \textbf{do}
\State \hspace*{0.4 cm} choose an index $i$ randomly from the set $S$
\State \hspace*{0.4 cm} \textbf{if} $(k \neq i )$ \textbf{then}
\State \hspace*{0.6 cm}  $x_k^{(t+1)} \sim p( x_k | x_1^{(t+1)},..,x_{k-1}^{(t+1)} , x_{k+1}^{(t)},..,x_{NM}^{(t)},\mathbf{y} , \mathbf{H} ) $
\State \hspace*{0.4 cm} \textbf{else}
\State \hspace{0.6 cm} generate the pmf $ p(x_k^{(t+1)} = j ) \sim U[0,1], \: \forall j \in \mathbb{A}$ 
\State \hspace{0.6 cm} From this pmf, sample $x_k^{(t+1)}$
\State \hspace*{0.4 cm} \textbf{end if}
\State \hspace*{0.2 cm} \textbf{end for}
\State \hspace*{0.2 cm} $\zeta = f_{{\tiny \mbox{ML}}} (\mathbf{x}^{(t+1)})$
\State \hspace*{0.2 cm} \textbf{if} $( \zeta \leq \beta )$ \textbf{then}
\State \hspace*{0.4 cm} $\textbf{z} = \textbf{x}^{(t+1)}$ and $\beta = \zeta$
\State \hspace*{0.2 cm} \textbf{end if}
\State \hspace*{0.2 cm} $t=t+1$
\State \textbf{end while}
\State \textbf{Output} $\mathbf{z}$, the solution vector.
\end{algorithmic}
\label{alg2}
\end{algorithm}

Randomized Gibbs sampling algorithm involves a randomization in the 
update rule. Instead of using the update rule as in conventional Gibbs 
sampling  with probability 1, a randomized rule using a parameter 
$r=\frac{1}{NM}$ is used as follows: with probability $(1-r)$, use 
conventional Gibbs sampling, and with probability $r$, take samples 
from a uniform distribution. That is, generate $|\mathbb{A}|$ 
probability values from a uniform distribution given by
\begin{equation}
p(x_i^{(t+1)} = i) \sim U[0,1], \: \forall i\in \mathbb{A},
\end{equation}
such that $\sum_{i=1}^{|\mathbb{A}|} p (x_i^{(t+1)}=i)=1$, and sample 
$x_k^{(t+1)}$ from this pmf. A listing of the randomized Gibbs sampling 
based detection algorithm is given in Algorithm \ref{alg2}.

\subsection{ Performance results}
\label{sec4c}
In this subsection, we present the BER performance of OTFS using the
randomized Gibbs sampling algorithm based signal detection. Perfect 
channel knowledge is assumed at the receiver. The number of signal 
propagation paths (taps) $P$ is taken to be 5. The Doppler model used 
is given by \cite{otfs5} 
\begin{equation}
\nu_i = \nu_{max} \cos \theta_i,
\end{equation}
where $\nu_i$ is the Doppler shift due to the $i$th path and $\theta_i$
is uniformly distributed, i.e., $\theta_i \sim U(0,\pi)$. The 5-tap delay 
model in Table \ref{delay_profile} with $T_s=2.1 \mu$s spacing with a 
uniform power delay profile is used.
\begin{table}[h]
\begin{center}
\begin{tabular}{ |c|c|c|c|c|c| } 
\hline
Path index $(i)$ & 1 &  2 &  3 &  4 &  5  \\ 
\hline
Excess Delay $(\tau_i)$ & $0 \mu s$ & $2.1 \mu s$ & $4.2 \mu s$  & $6.3 \mu s$  & $8.4 \mu s$   \\
\hline
\end{tabular}
\caption{5-tap delay model with uniform delay profile.}
\label{delay_profile}
\vspace{-6mm}
\end{center}
\end{table}

\begin{table}[t]
\begin{center}
\begin{tabular}{|p{0.50\linewidth}|p{0.25\linewidth}|}
\hline
\textbf{Parameter} & \textbf{Value} \\
\hline
Carrier frequency (GHz) & 4 \\
\hline
Subcarrier spacing (kHz) &  3.75\\
\hline
Frame size $(M,N)$ & $(128,32)$ \\
\hline
Modulation scheme & BPSK \\
\hline
UE speed (kmph) & 27, 120, 500\\
\hline
Channel knowledge & Perfect \\
\hline
\end{tabular} 
\vspace{2mm}
\caption{Simulation parameters.}
\label{SimPar}
\vspace{-10mm}
\end{center}
\end{table}

The received signal model in (\ref{inpopsparse}) is used for the simulation
with the parameter $E_i =4$. Three different user equipment (UE) speeds, 
viz., 27 kmph, 100 kmph, and 500 kmph are considered. At a carrier frequency
of 4 GHz, these speeds correspond to Doppler frequencies 100 Hz, 444.44 Hz, 
and 1851 Hz, respectively. Other simulation parameters used are given in 
Table \ref{SimPar}. 

\begin{figure}
\centering
\includegraphics[width=9.5 cm, height= 7.5cm]{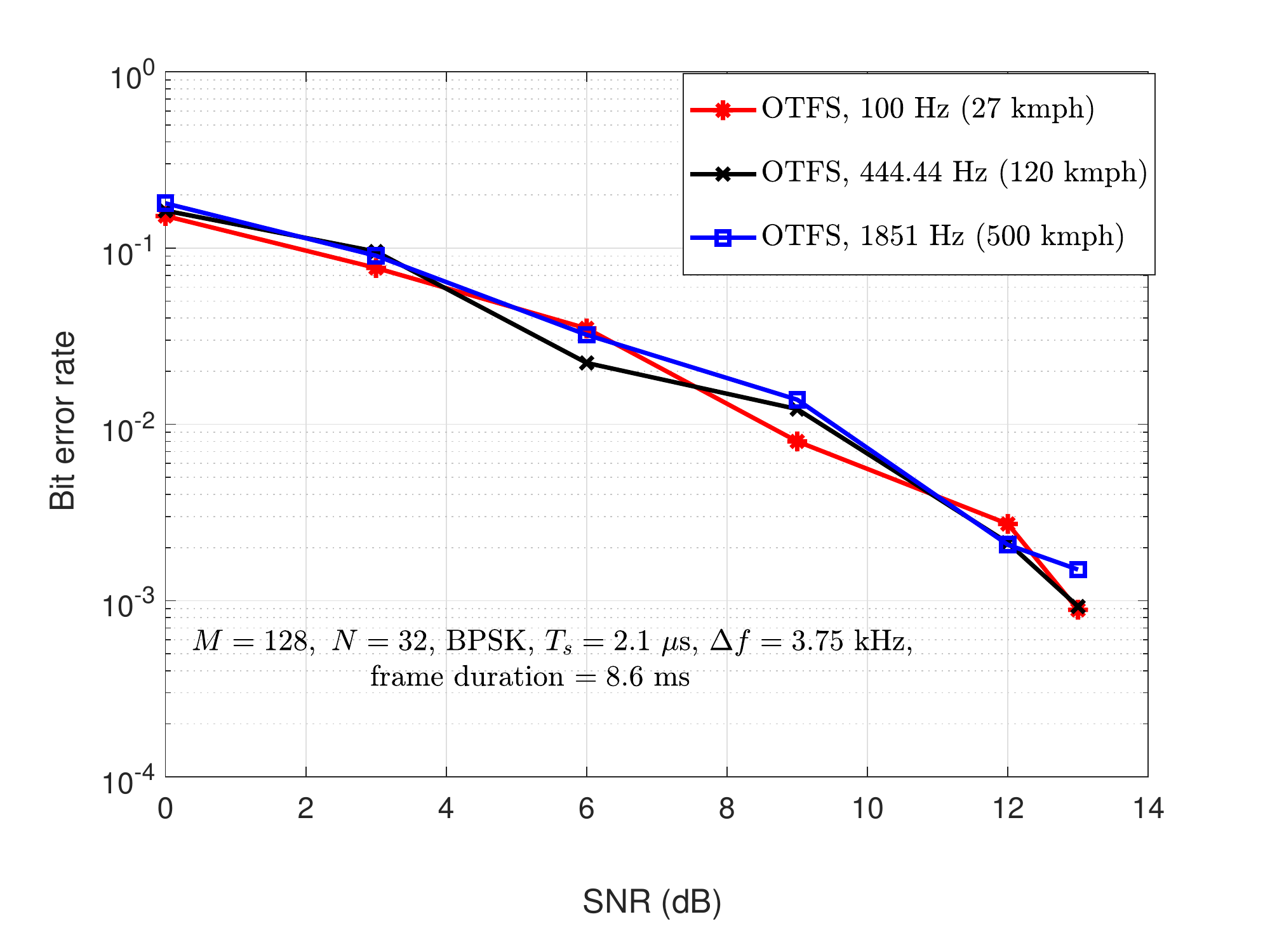}
\vspace{-7mm}
\caption{BER performance of OTFS using randomized Gibbs sampling detection 
for 3 different Dopplers 100 Hz, 444.44 Hz, and 1851 Hz.}
\label{BER2}
\vspace{-4mm}
\end{figure}

Figure \ref{BER2} shows the BER performance of OTFS with $M=128$, $N=32$, 
BPSK, $\Delta f=3.75$ kHz subcarrier spacing, and a frame length of 
$MNT_s=128\times32\times2.1\times 10^{-6}=8.6$ msec, using randomized 
Gibbs sampling based detection algorithm. The number of iterations used 
is 3. From Fig. \ref{BER2}, we observe that a BER of $10^{-3}$ is 
achieved at an SNR value of about 13 dB for all the three Doppler 
frequencies considered. Note that the Doppler spread for 500 kmph UE 
speed at 4 GHz carrier frequency is 1.851 kHz, which in conventional 
systems like OFDM would cause severe ICI and performance degradation 
for the considered subcarrier spacing of $3.75$ kHz. In fact, OFDM is 
known to breakdown completely at these high Dopplers. Whereas, the 
the signal localization achieved by OTFS in the delay-Doppler domain 
renders good BER performance which is almost invariant to Doppler. The 
BER plots in Fig. \ref{BER2}, therefore, give a clear illustration of 
the performance robustness of OTFS in high Doppler fading channels. 

\section{Delay-Doppler Channel Estimation}
\label{sec5}
In the previous section, we assumed perfect channel knowledge, i.e., 
$\mathbf{H}$ is assumed to be perfectly known at the receiver. Here, we 
relax this assumption and present a method for estimating the channel in 
the delay-Doppler domain. The method uses a PN sequence as pilot for 
channel estimation. The estimation approach is as follows.
The estimation is done in the discrete domain, where three quantities
of interest, namely, delay shift $(\delta_i)$, Doppler shift $(\nu_i)$, 
and fade coefficient $(\alpha_i)$ for the $i$th path, for all $i=1,\cdots,P$,
need to be estimated. The estimation of $\delta_i$ and $\nu_i$ is first 
done by solving a time-frequency shift problem which involves the
computation of a matched filter matrix (described later in Sec. 
\ref{sec_new1}) for each $i$, and the estimates of $\alpha_i$s are 
obtained to be the values of $P$ highest peaks in the matched filter 
matrix. 

Consider the delay-Doppler channel impulse response $h(\tau,\nu)$ in 
(\ref{sparsechannel}). The coupling between the input signal and the 
channel can be written as
\begin{equation}
\label{finitechannelcoupling}
y(t)=\sum_{i=1}^{P} h_i x(t-\tau_i)e^{j2\pi \nu_i (t-\tau_i)}+v(t).
\end{equation}
Let $ \mathcal{H} $ denote the vector space of complex valued functions 
on the set of finite integers $\mathbb{Z}_{N_p} =\{0,1,\cdots,N_p-1\}$ 
equipped with addition and multiplication modulo $N_p$. The inner product 
in $\mathcal{H}$ is defined as 
\begin{equation}
\langle f_1,f_2 \rangle=\sum\limits_{{n \in \mathbb{Z}_{N_p}}}f_1[n]f^*_2[n],
\ \ f_1,f_2 \in \mathcal{H}.
\end{equation}
Also, define $e(t) = e^{{j2\pi\over N_p}t}$.
 
\subsection{Channel parameters and the discrete channel model}
\label{sec5a}
Here, we present the discrete channel model for (\ref{finitechannelcoupling}), 
which simplifies the problem of channel estimation (since waveforms becomes 
sequences). The model in (\ref{finitechannelcoupling}) can be written in a 
slightly modified form as
\begin{equation}
y(t) = \sum_{i=1}^{P} h_i' e^{j2\pi \nu_i t}x(t-\tau_i) + v(t),
\end{equation}
where $h_i'=h_i e^{-j2\pi \nu_i \tau_i}$, 
$\nu_i \in \mathbb R,\ \tau_i \in \mathbb R_+ $. The parameters 
$(h_i', \nu_i, \tau_i)$ for $i=1,2,\cdots,P,$ are called the channel 
parameters and we have to estimate them. They, in turn, give the 
estimate of $\mathbf{H}$ in (\ref{vecform}). The process for converting 
the continuous time channel model to a discrete channel model is 
described as follows \cite{channelestimation1}. Start with a sequence 
$S \in \mathcal{H}$ and transmit the following analog signal
\begin{equation*}
S_A(t) = \sum_{n=0}^{M-1} S[n\: \mbox{mod}\: N_p]\mbox{sinc}(Wt-n),
\end{equation*} 
where $M \geq N_p$. Let $T_{\mbox{\tiny{spread}}}=\mbox{max}(\tau_i)$ 
denote the time spread of the channel and define 
$K= \lceil WT_{\mbox{\tiny{spread}}} \rceil$. Let $M=N_p+K$. Transmit 
the analog signal $S_A(t)$ from time $t=0$ to $t = {M \over W}$. The 
received signal $R_A(t)$ is related to $S_A(t)$ through 
(\ref{finitechannelcoupling}). $R_A(t)$ is sampled at an interval 
$T_s={1 \over W}$ from time $K\over W$, and the sequence 
$R[n]=R_A({(K+n) \over W})$ for $n=0,1,\cdots,N_p-1$ is obtained. The 
following proposition in \cite{channelestimation1} gives the discrete 
channel model.
\begin{prop}
\label{prop1}
Let $\tau_i \in {1 \over W}\mathbb{Z}_{+}$ and 
$\nu_i \in {W \over N_p}\mathbb{Z}$ for $i=1,2,\cdots,P$. 
Then $R[n]$ given above satisfies 
\begin{equation*}
R[n] = \sum_{i=1}^{P}\alpha_i e(\omega_i n)S[K+n-\delta_i] + v[n], \:\: n \in \mathbb{Z}_{N_p},
\end{equation*} 
where $\alpha_i= h_i'e^{(j2\pi\nu_iK/W)},\  \delta_i = \tau_i W$, and $\omega_i = N_p \nu_i/W$. 
\end{prop}

\subsection{PN pilot based channel estimation}
\label{sec5b}
Let $S,R \in \mathcal{H}$, where $R[n]$ is given by
\begin{equation}
\label{discreteMF}
R[n] = \sum_{i=1}^{P}\alpha_i e(\omega_i n)S[n-\delta_i] +v[n], \:\: n \in \mathbb{Z}_{N_p},
\end{equation}
where $\alpha_i \in \mathbb{C}$, $\delta_i,\omega_i \in \mathbb{Z}_{N_p}$, 
and $v[n] \in \mathcal{H}$. Once $\alpha_i, \delta_i$, and $\omega_i$ 
are estimated, we can compute $h_i',\  \tau_i$, and  $\nu_i$ for 
$i=1,2,\cdots,P$ using Proposition \ref{prop1}, which solves the channel 
estimation problem. We estimate $\delta_i$, $\omega_i$, and $\alpha_i$ 
using a PN pilot based scheme described as follows.

\begin{figure*}
\centering
\subfigure[ Auto-correlation ]
{\includegraphics[height=5.5cm,width=6cm]{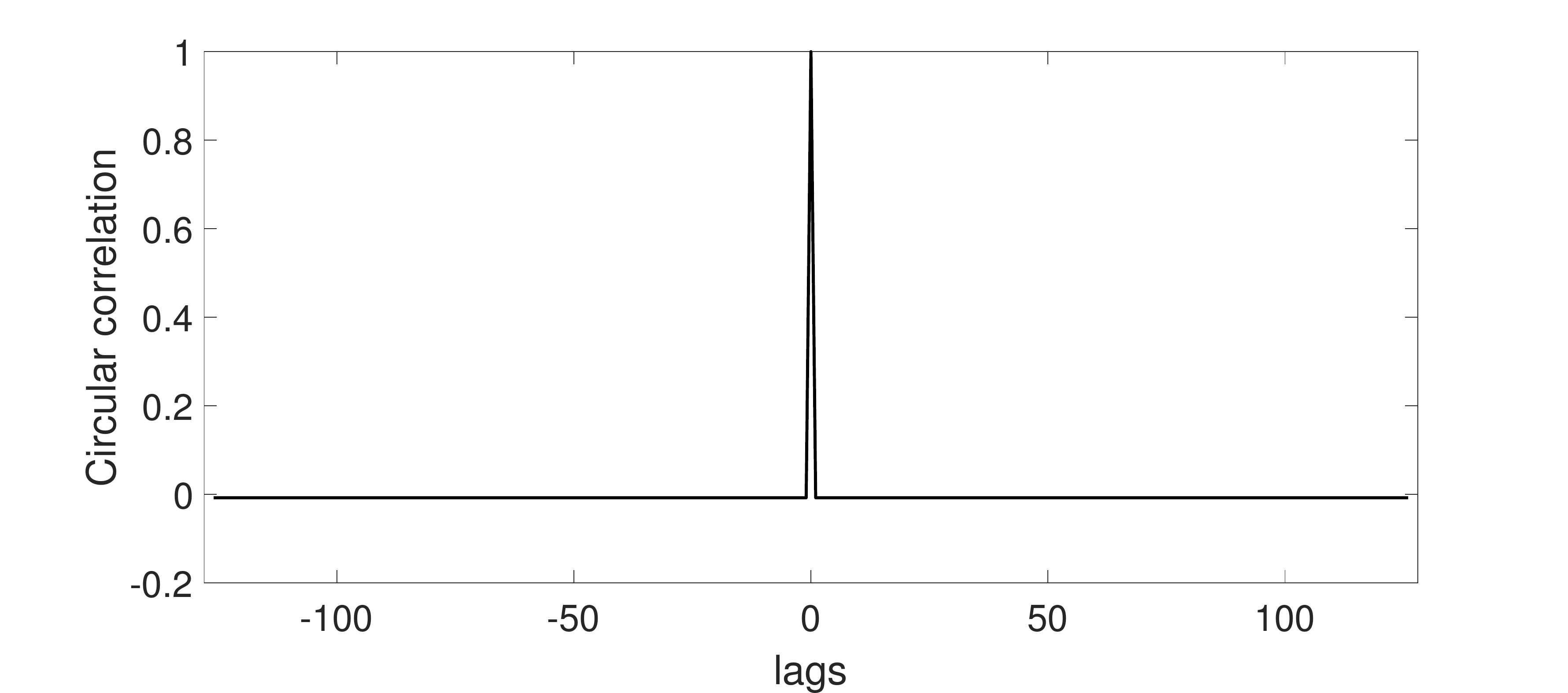}
\label{ACF}
}
\hspace{-4mm}
\subfigure [ ( $\delta_0, \omega_0 $) =  (40,90)  ]
{\includegraphics[height=5.5cm,width=6cm]{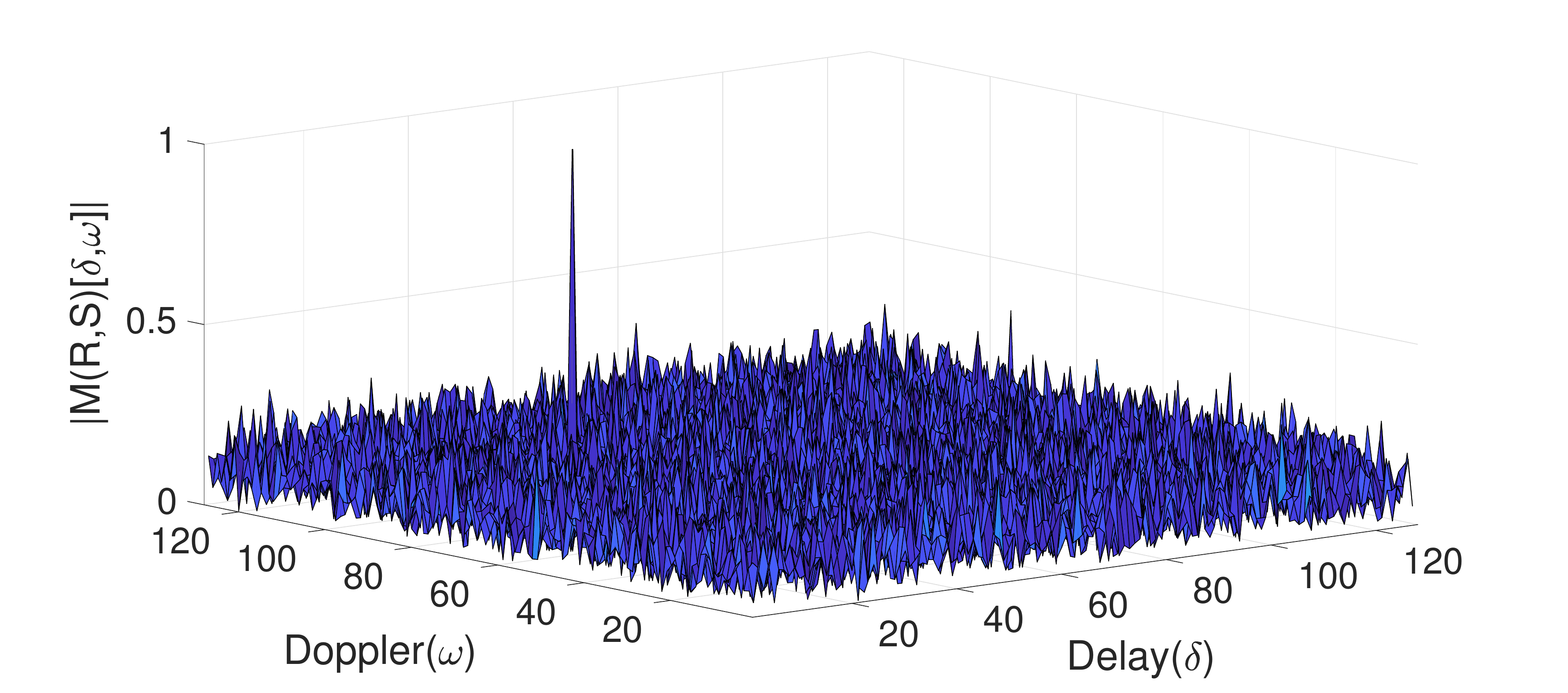}
\label{TFS1}
}
\hspace{-4mm}
\subfigure[ ( $\delta_0, \omega_0 $) =  (80,60) ]
{\includegraphics[height=5.5cm,width=6cm]{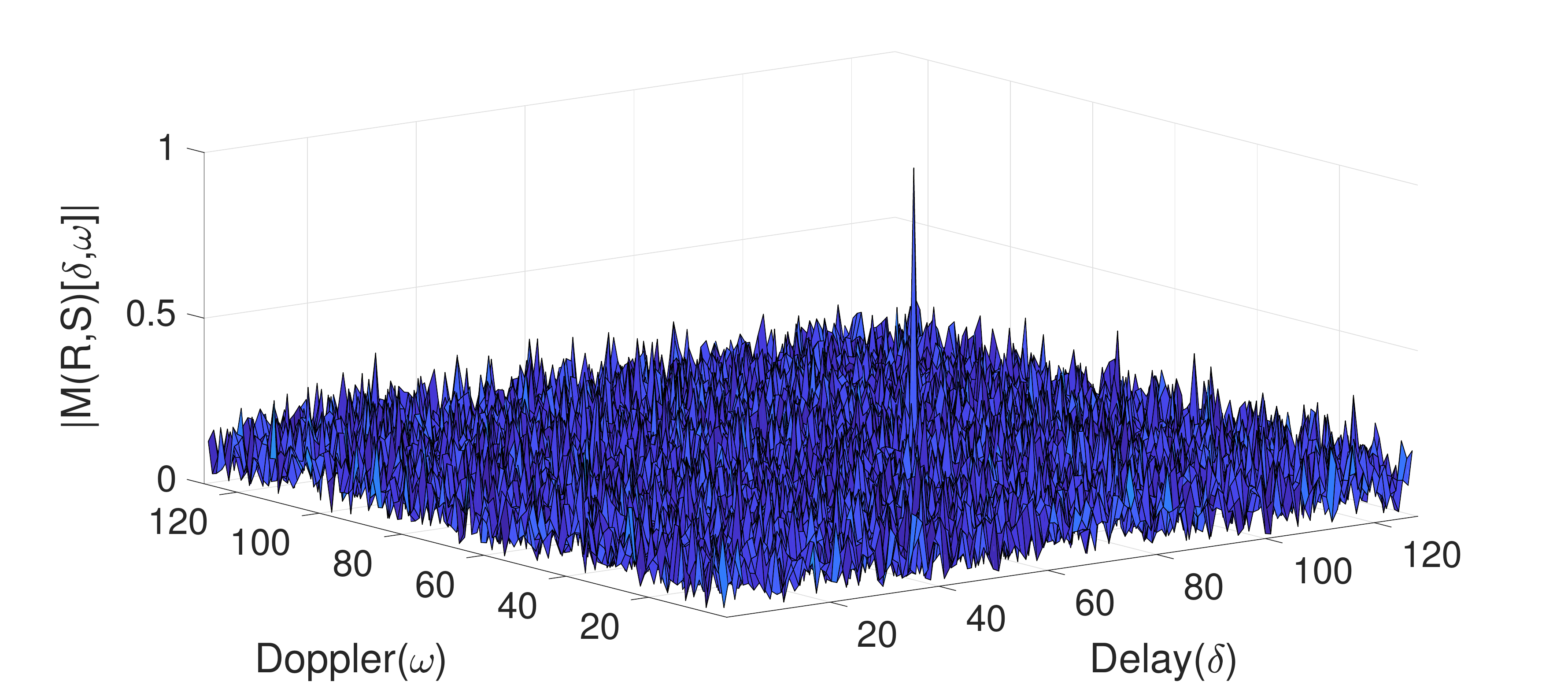}
\label{TFS2}
}
\vspace{-1mm}
\caption{(a) Auto-correlation function of the PN sequence with 
$N_p = 127$. (b), (c) Magnitude of the entries of the matched filter 
matrix at SNR = 0 dB and $N_p = 127$ for the time-frequency shift problem 
with $(\delta_0,\omega_0)=(40,90)$ and $(\delta_0,\omega_0)=(80,60)$, 
respectively.}
\vspace{-3mm}
\label{Ch_Est1}
\end{figure*}

\subsubsection{Solving for $(\delta_i,\omega_i,\alpha_i)$}
\label{sec_new1}
Consider the following simpler variant of (\ref{discreteMF}):
\begin{equation}
\label{TFS}
R[n] = e(\omega_{0}n)S[n-\delta_{0}] + v[n], 
\end{equation}
where $n \in \mathbb{Z}_{N_p}$ and 
$(\delta_0,\omega_0) \in \mathbb{Z}_{N_p} \times \mathbb{Z}_{N_p} $. 
The problem of estimating  $(\delta_0,\omega_0)$ is called the 
time-frequency shift problem, which is solved as follows. Define
the matched filter matrix of $R$ and $S$ as
\begin{equation}
\mathcal{M}(R,S)[\delta,\omega] = \langle R[n], e(\omega n)S[n-\delta] \rangle,\:\: (\delta,\omega) \in \mathbb{Z}_{N_p} \times \mathbb{Z}_{N_p}.
\end{equation}
Suppose $R$ and $S$ satisfy (\ref{TFS}) and $S \in \mathcal{H}$ is a PN 
sequence of norm one and length $N_p$. Then, we can obtain the following
expression for $\mathcal{M}(R,S)[\delta,\omega]$ by the law of iterated 
logarithm, with probability going to one, as $N_p$ goes to infinity
\cite{channelestimation1}:
\begin{equation}
\label{MFA}
\begin{split}
\mathcal{M}(R,S)[\delta,\omega] &= \ 1+\epsilon_{N_p}' \: \: \mbox{if} \: \: (\delta,\omega) = (\delta_0,\omega_0)\\
& = \ \epsilon_{N_p} \: \: \hspace{0.60 cm} \mbox{if} \: \: (\delta,\omega) \neq (\delta_0,\omega_0),  \\ 
\end{split}
\end{equation}
where $|\epsilon_{N_p}'| \leq {1 \over \sqrt{N_p}}$ and 
$|\epsilon_{N_p}| \leq {(C+1) \over \sqrt{N_p}}$ for some constant 
$C > 0$. Hence, the solution to the time-frequency shift problem is 
to compute $\mathcal{M}(R,S)$ and choose $(\delta_0,\omega_0)$ for 
which $\mathcal{M}(R,S)[\delta_0,\omega_0] \approx 1$. Once 
$(\delta_i,\omega_i)$'s are obtained as above, the identity 
(\ref{MFA}) along with the bi-linearity of the inner product gives
\begin{equation}
\label{MFAA}
\alpha_i \approx \mathcal{M}(R,S)[\delta_i,\omega_i] , \: i=1,2,\cdots,P.
\end{equation}
This solves the problem of estimating $(\alpha_i,\delta_i,\omega_i)$. 

{\em Example 1:} For illustration, we present plots of the PN 
sequence ACF and the magnitudes of the entries of the matched filter 
matrix in Fig. \ref{Ch_Est1}. Figure \ref{ACF} shows the ACF of the 
PN sequence with $N_p=127$, and Figs. \ref{TFS1} and \ref{TFS2} show 
the magnitudes of the entries of the matched filter matrix for the 
time-frequency shift problem with $\alpha_0 =1$ at SNR = 0 dB for 
$(\delta_0,\omega_0)=(40,90)$ and $(\delta_0,\omega_0)=(80,60)$, 
respectively. From these figures, we see that the magnitude of the 
matched filter matrix is close to one 
when $(\delta,\omega)=(\delta_0,\omega_0)$ and is close to zero when 
$(\delta,\omega) \neq (\delta_0,\omega_0)$, as indicated by (\ref{MFA}). 

\begin{figure*}
\centering
\subfigure[ $(\delta_0, \omega_0,\delta_1, \omega_1) =  (100, 30, 10, 80)$ ]
{\includegraphics[height=5.5cm,width=9cm]{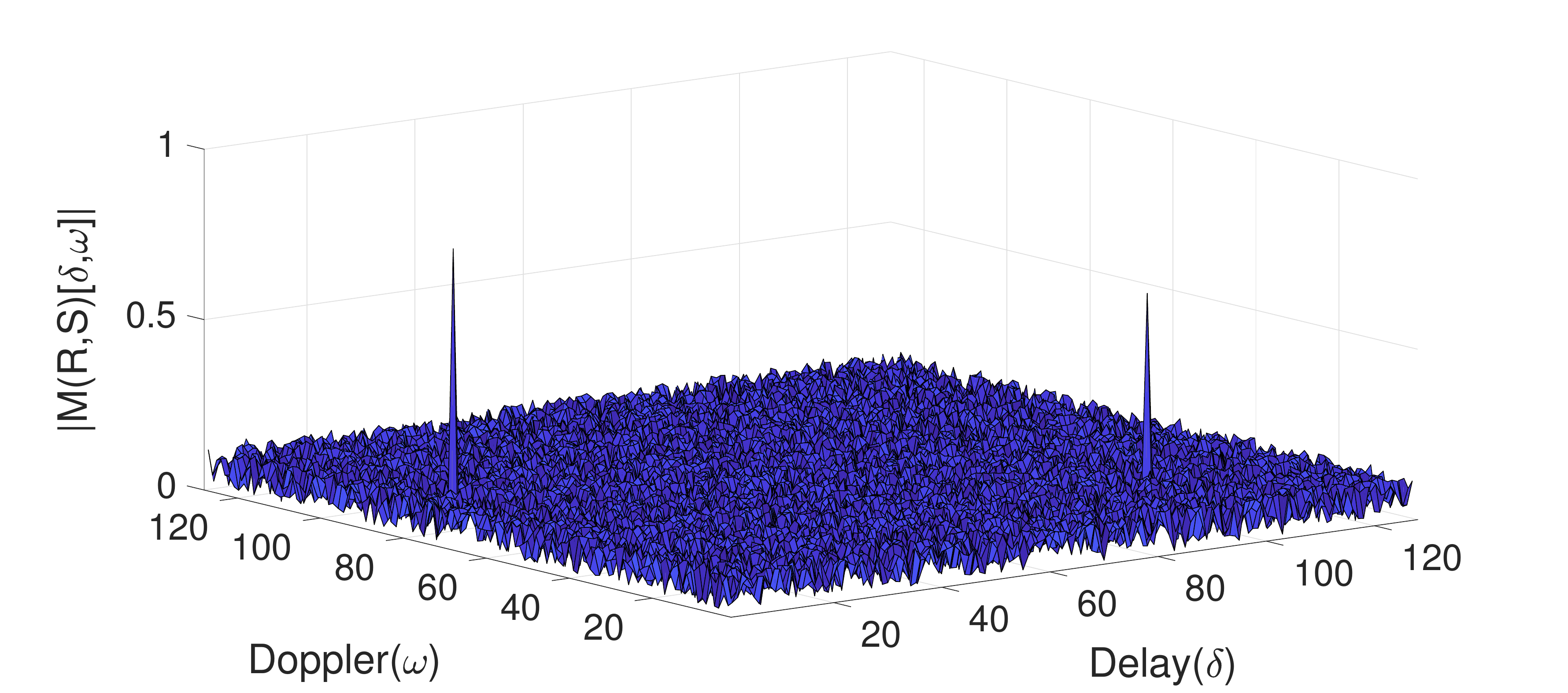}
\label{2paths1}
}
\hspace{-4mm}
\subfigure[ $(\delta_0, \omega_0, \delta_1, \omega_1) = (60, 20, 70, 70)$ ]
{\includegraphics[height=5.5cm,width=9cm]{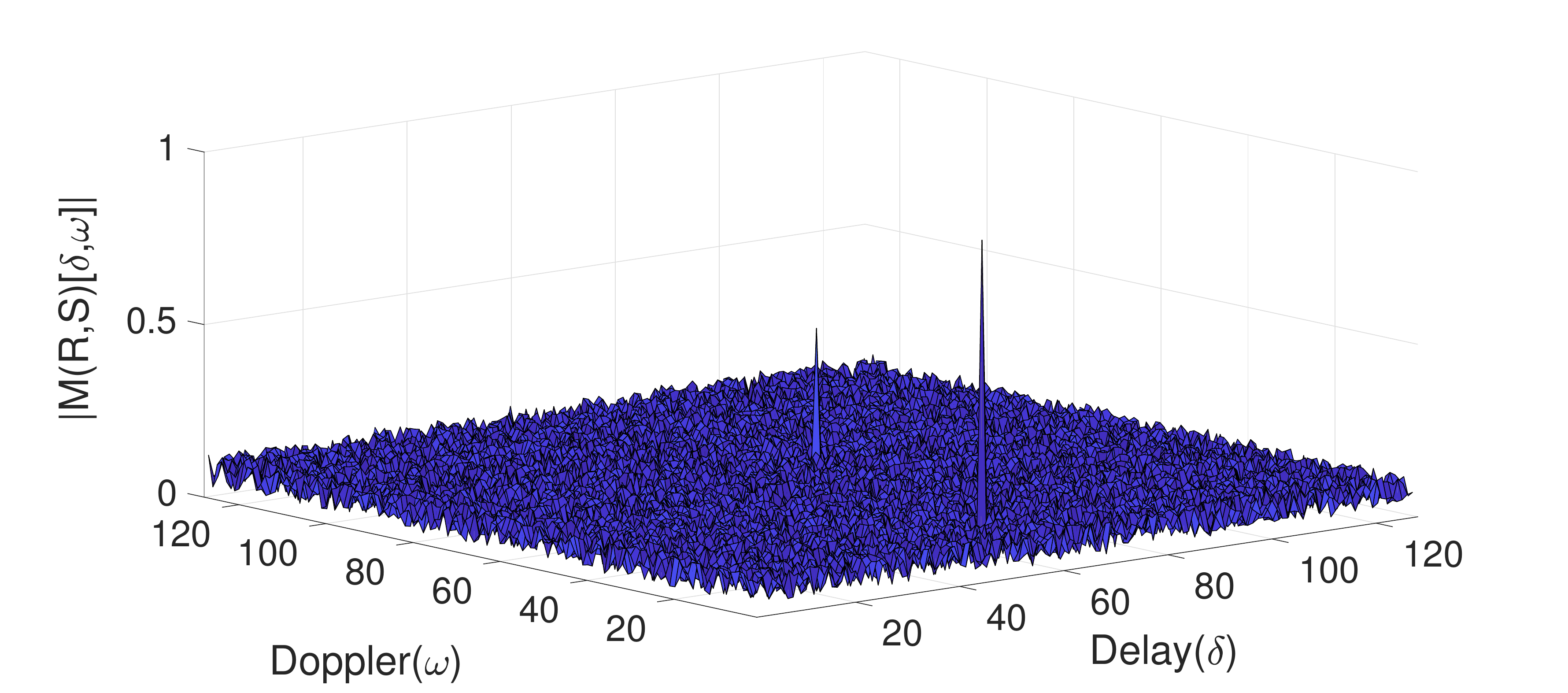}
\label{2paths2}
}
\vspace{-1mm}
\caption{Magnitude of the entries of the matched filter matrix at 
SNR = 20 dB and $N_p = 127$ for channels with $P=2$. (a)
$(\delta_0, \omega_0,\delta_1, \omega_1) =  (100, 30, 10, 80)$,
(b) $(\delta_0, \omega_0, \delta_1, \omega_1) = (60, 20, 70, 70)$. }
\vspace{-2mm}
\label{Ch_Est2}
\end{figure*}

\begin{figure*}
\centering
\subfigure[$N_p=127$ ]
{\includegraphics[height=5.5cm,width=9cm]{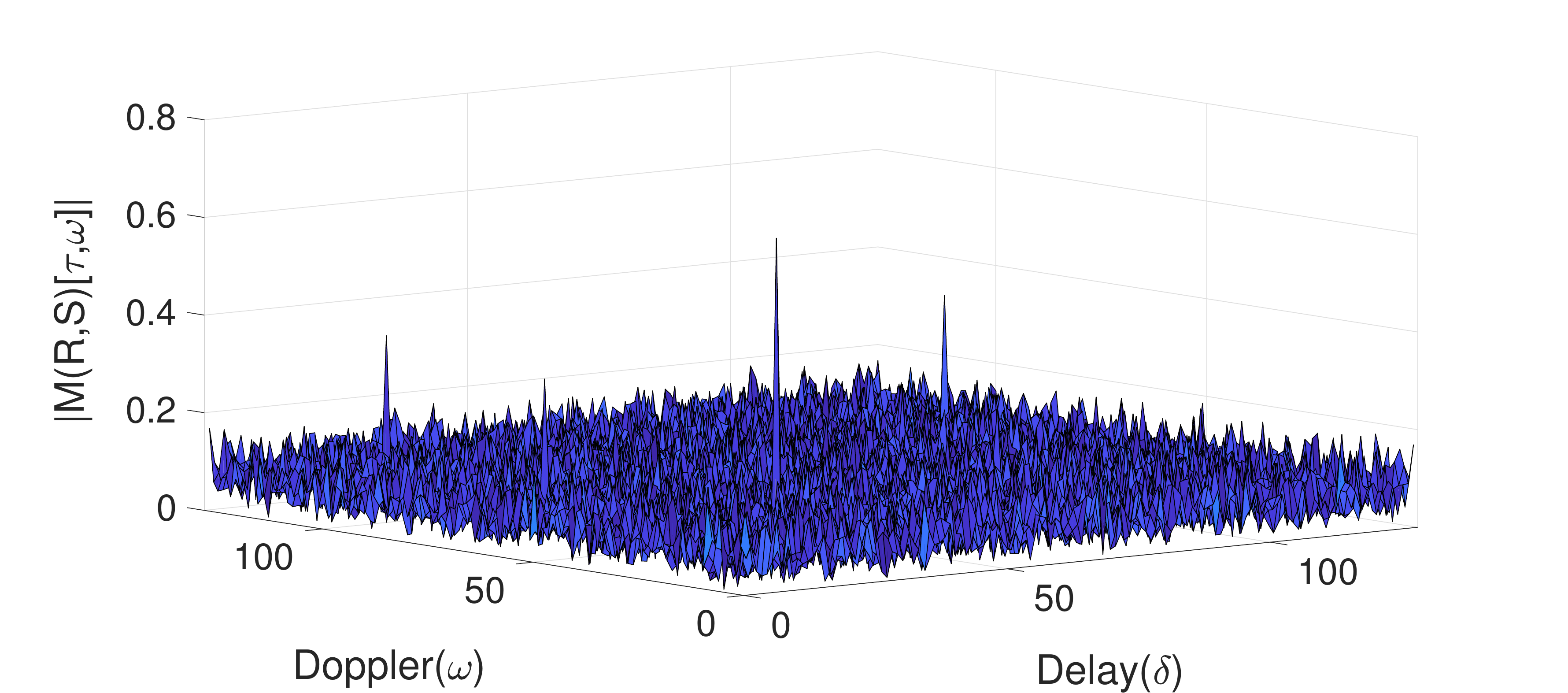}
\label{5paths1}
}
\hspace{-4mm}
\subfigure[$N_p=1023$ ]
{\includegraphics[height=5.5cm,width=9cm]{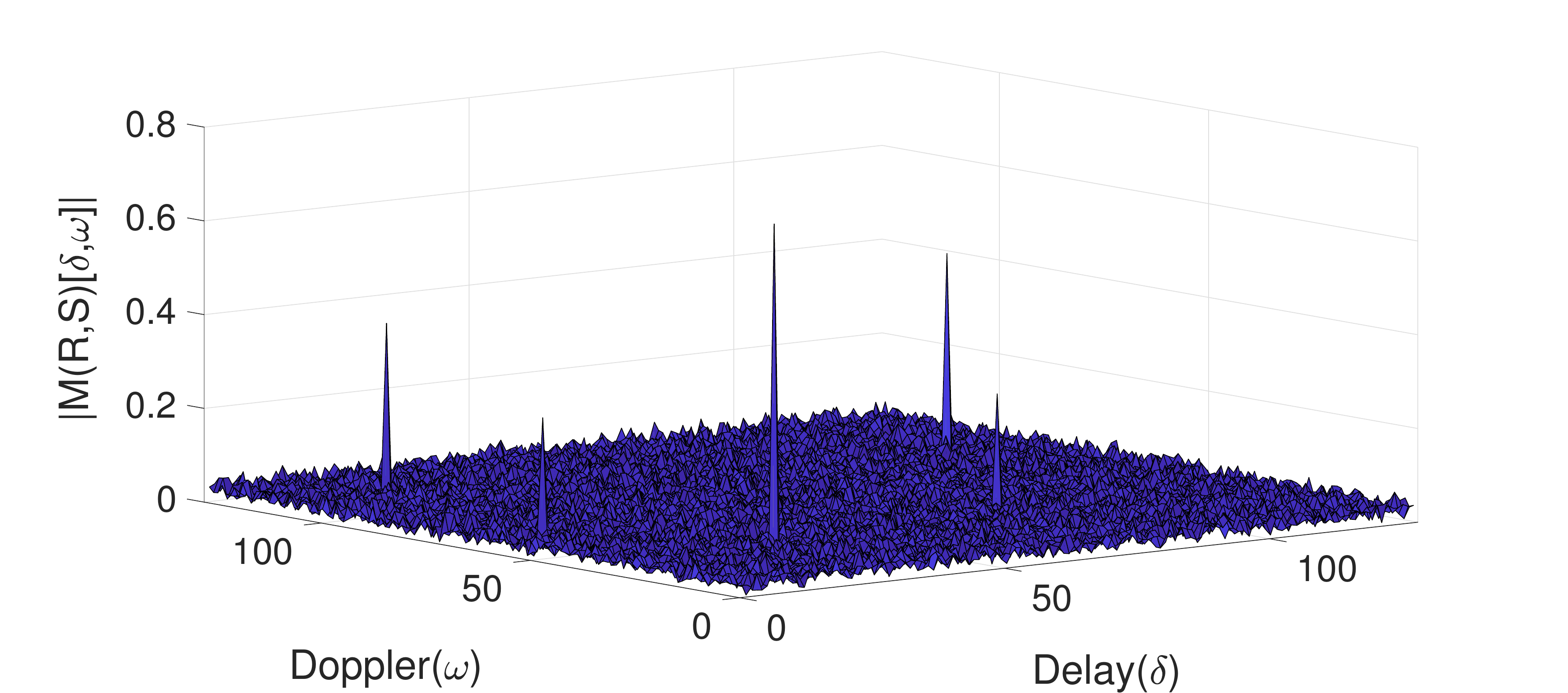}
\label{5paths2}
}
\vspace{-1mm}
\caption{Magnitude of the entries of the matched filter matrix at 
SNR = 20 dB for a channel with $P=5$, 
$(\delta_i,\omega_i) \in \{(10,60),(20,110),(30,30),(80,40),(110,90)\}$ 
for (a) $N_p = 127$ and (b) $N_p = 1023$.}
\vspace{-2mm}
\label{Ch_Est3}
\end{figure*}

{\em Example 2:}
The magnitude plots of the entries of the matched filter matrix for 
channels with $P=2$ at SNR = 20 dB are shown in Fig. \ref{2paths1} 
and Fig. \ref{2paths2}, respectively, for
$(\delta_0, \omega_0,\delta_1, \omega_1) =  (100, 30, 10, 80)$ and
$(\delta_0, \omega_0, \delta_1, \omega_1) = (60, 20, 70, 70)$.
We can observe that there are two strong peaks at 
$(\delta,\omega)=(\delta_i,\omega_i)$, $i=0,1$. The entries in the
matched filter matrix corresponding to these two peaks give 
the corresponding $\alpha_i$ values. The values of 
the magnitudes are very small when $(\delta,\omega)\neq(\delta_i,\omega_i)$.

{\em Example 3:}
In Fig. \ref{Ch_Est3}, we show the effect of $N_p$ on the estimation 
accuracy. This is illustrated using the magnitude plots for a channel 
with $P=5$ at SNR = 20 dB and  
$(\delta_i,\omega_i) \in \{(10,60),(20,110),(30,30),(80,40),(110,90)\}$.
Two values of $N_p$ are considered. Figure \ref{5paths1} considers
$N_p=127$ and Fig. \ref{5paths2} considers $N_p=1023$. From these two
figures, we can see that the magnitudes for 
$(\delta,\omega)\neq(\delta_i,\omega_i)$ for $N_p=1023$ are low compared 
to those for $N_p=127$. That is, the estimate can be more accurate for 
larger values of $N_p$, which can be seen from (\ref{MFA}) and the fact 
that $|\epsilon_{N_p}|, |\epsilon_{N_p}'| \propto 1/ \sqrt{N_p}$. 

\subsection{Performance results}
\label{sec5c}
In this subsection, we present the BER performance of OTFS with estimated 
channel, where the channel parameters (and hence \textbf{H}) are estimated 
using the method described in the previous subsection. The worst case 
delay and Doppler spread are assumed to be known, so that the parameter 
$K$ can be chosen a priori in the process of continuous time to discrete 
time conversion. The channel model in (\ref{inpopnofracdopp}) is used, 
where the delay and Doppler values are assumed to be integer multiples. 
A carrier frequency of 4 GHz and a channel model with $P=5$ are considered. 
The delay-Doppler profile considered in the simulation is shown in Table 
\ref{tab_newx}. 
A frame size of $(M,N)=(32,32)$, subcarrier spacing of $\Delta f=15$ kHz, 
uniform power delay profile, and BPSK are considered.
\begin{table}[h]
\begin{center}
\begin{tabular}{ |c|c|c|c|c|c| } 
\hline
Path index $(i)$ & $1$ &  $2$ &  $3$ &  $4$ &  $5$  \\ 
\hline
Delay ($\tau_i$), $\mu$s & $2.1$  & $4.2 $  & $6.3$  & $8.4$  & $10.4$\\
\hline
Doppler ($\nu_i$), Hz & $0$ & $470$  & $940$  & $1410$ & $1880$   \\
\hline
\end{tabular}
\caption{Delay-Doppler profile for the channel model with $P=5$.}
\label{tab_newx}
\end{center}
\end{table}

\begin{figure*}
\centering
\subfigure[]
{\includegraphics[height=6.25cm,width=9cm]{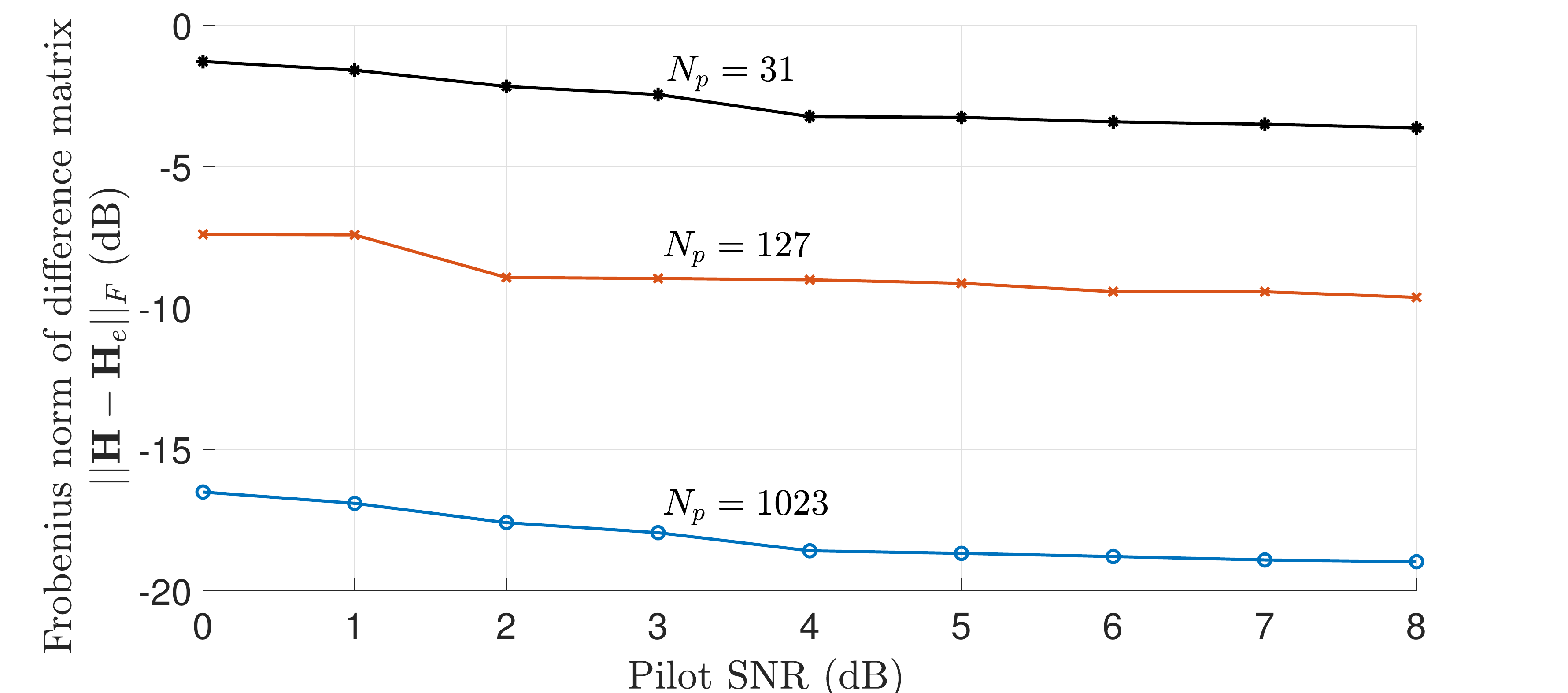}
\label{esterr_pilotsnr}
}
\hspace{-4mm}
\subfigure[]
{\includegraphics[height=6.25cm,width=9cm]{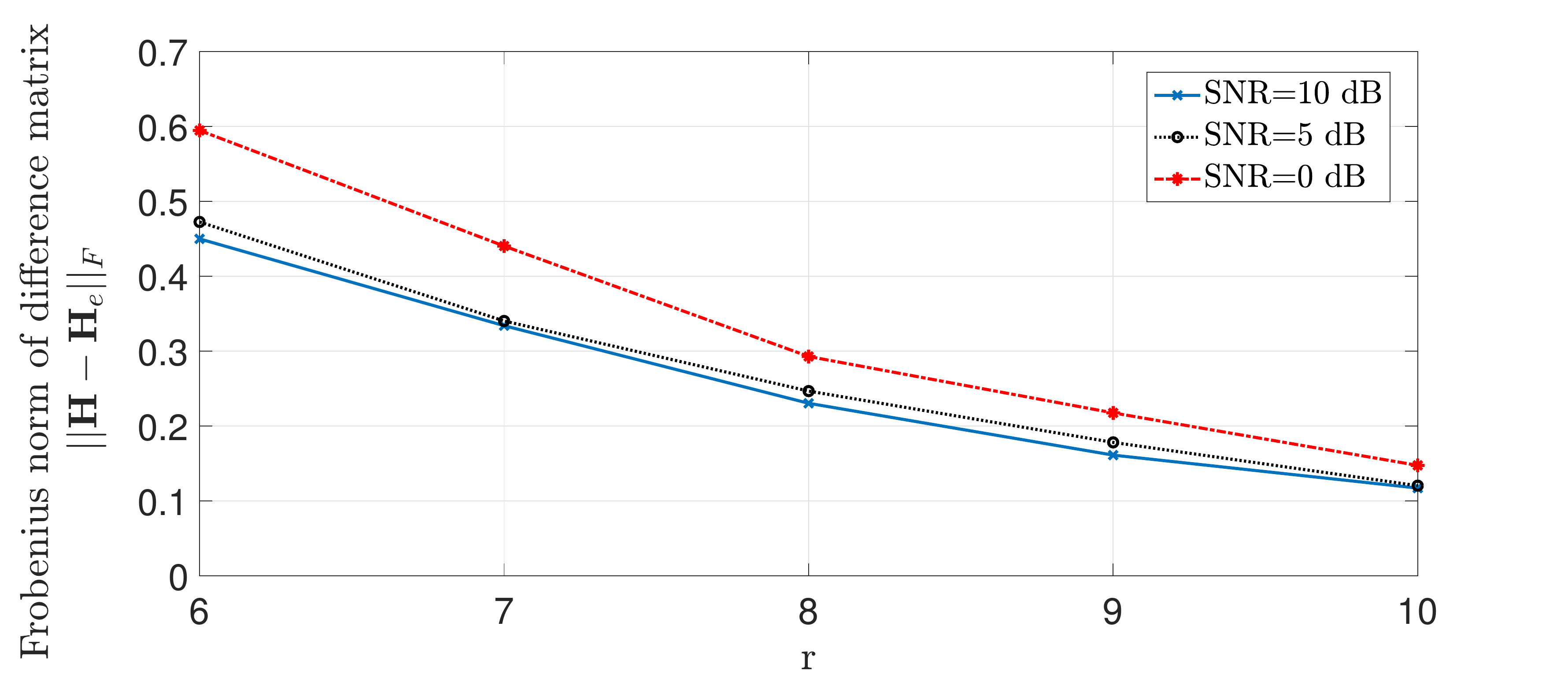}
\label{esterr_pnlength}
}
\vspace{-1mm}
\caption{Frobenius norm of the difference between the channel matrix 
($\textbf{H}$) and the estimated channel matrix ($\textbf{H}_e$) as a 
function of (a) pilot SNR for various $N_p$ values, and (b) $r$, where
$N_p = 2^{r}-1$, for various pilot SNR values.}
\vspace{-2mm}
\label{Ch_Est2}
\end{figure*}

In Fig. \ref{Ch_Est2}, we illustrate the accuracy of the proposed channel
estimation in terms of the estimation error given by the Frobenius norm of 
the difference between the channel matrix ($\textbf{H}$) and the estimated 
channel matrix ($\textbf{H}_e$), i.e., $\|{\bf H}-{\bf H}_e\|_F$. Figure
\ref{esterr_pilotsnr} shows the variation of $\|{\bf H}-{\bf H}_e\|_F$ as 
a function of pilot SNR for three different values of $N_p$ ($=31,127,1023$). 
Figure \ref{esterr_pnlength} shows the variation of $\|{\bf H}-{\bf H}_e\|_F$
as a function of $r$, where $N_p=2^r-1$, for three different values of
SNR ($=0,5,10$ dB). From these figures, we see that the estimation error
decreases as SNR increases, which is expected. Also, larger PN sequence
lengths ($N_p$) reduce the estimation error because of the $1/\sqrt{N_p}$
relation between $N_p$ and the estimation error, as indicated by  
(\ref{MFA}). 

\begin{figure*}
\centering
\subfigure[]
{\includegraphics[height=6.75cm,width=9cm]{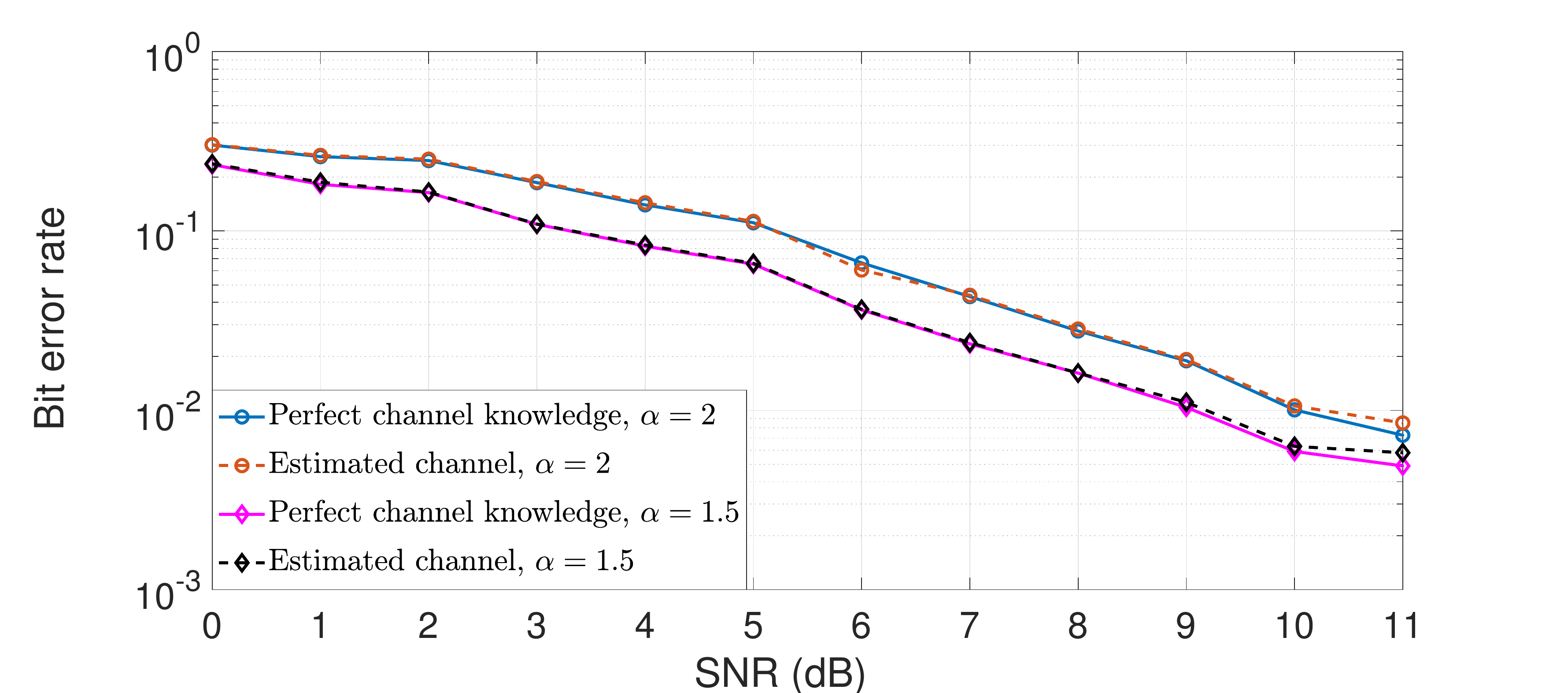}
\label{est_det1}
}
\hspace{-4mm}
\subfigure[]
{\includegraphics[height=6.75cm,width=9cm]{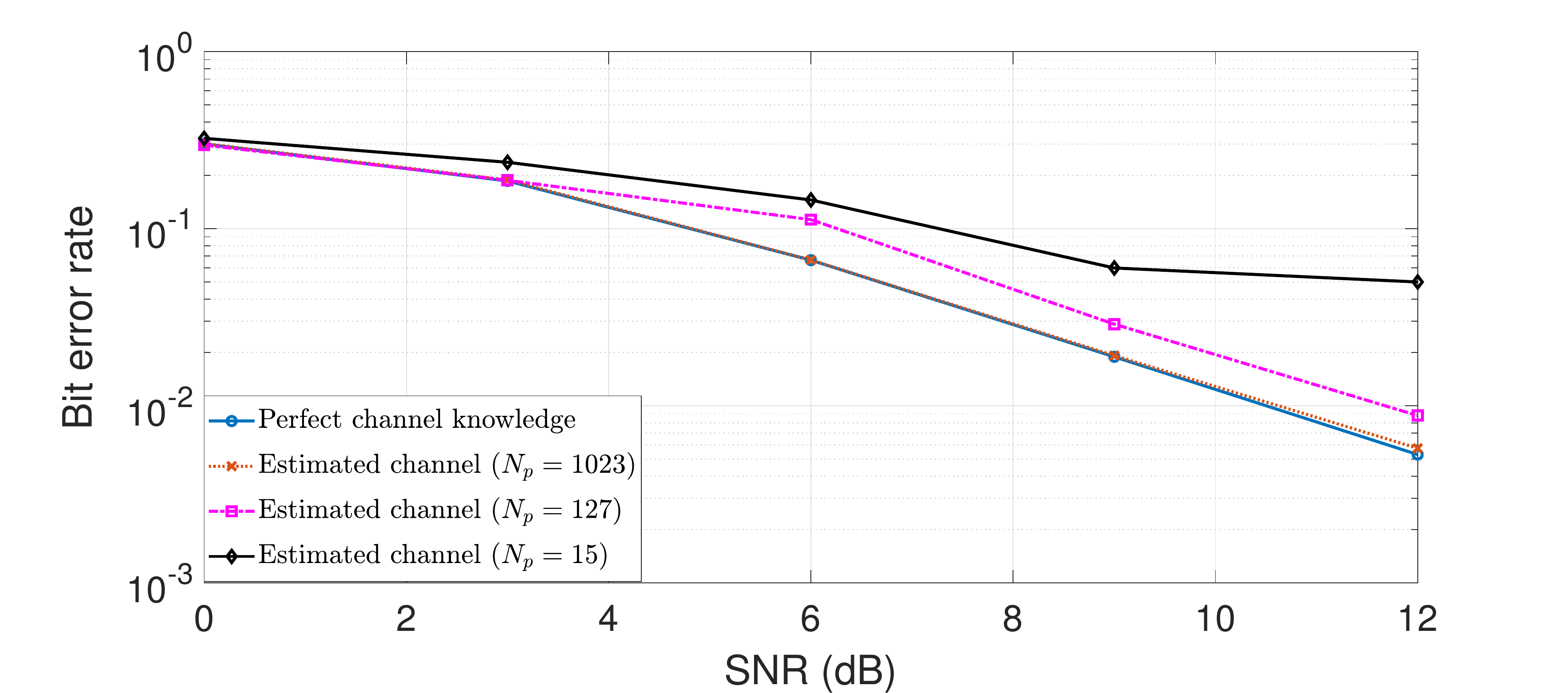}
\label{est_det2}
}
\vspace{-1mm}
\caption{BER performance of OTFS with estimated channel. (a) $N_p=1023$ and 
$\alpha=1.5,2$. (b) $\alpha=2$ and $N_p=15,127,1023$.}
\vspace{-2mm}
\label{Ch_Est2x}
\end{figure*}

In Fig. \ref{Ch_Est2x}, we present the BER performance of OTFS with
the estimated channel matrix. Detection using Gibbs sampling algorithm
with different temperature parameters are considered. Figures 
\ref{est_det1} and \ref{est_det2} show the BER versus SNR plots for 
the cases of perfect channel knowledge and estimated channel knowledge.
In Fig. \ref{est_det1}, BER performance when $N_p=1023$ and $\alpha=1.5,2$ 
are shown. In Fig. \ref{est_det2}, BER performance for 
$N_p=15,127,1023$ when $\alpha=2$ are shown. It is observed that the 
the performance degradation with estimated channel knowledge (relative
to the performance with perfect channel knowledge) is not significant
when $N_p=1023$, because of the low estimation error achieved with 
such large $N_p$. The degradation, however, becomes significant when
$N_p$ is reduced. For example, the performance degradation is about
1 dB at $10^{-2}$ BER when $N_p=127$, and the degradation gets severe
when $N_p=15$. While the estimates become more accurate when $N_p$ is 
increased, the estimation complexity also increases with increasing $N_p$. 
Thus, there is a trade-off between the accuracy of estimation and 
complexity for the choice of $N_p$.

\section{Conclusions}
\label{sec6}
We investigated OTFS modulation, which is a recently proposed modulation 
suited for communication in high-Doppler fading channels, from a signal 
detection and channel estimation perspective. In particular, we proposed
a low-complexity detection scheme based on MCMC sampling techniques and a 
PN pilot sequence based channel estimation scheme in the delay-Doppler 
domain. Our results showed that the BER performance of OTFS is robust 
even at high-Doppler frequencies (e.g., 100 Hz, 444 Hz, and 1851 Hz 
Dopplers).  The proposed channel estimation scheme was shown to achieve 
small estimation errors and BER degradation for large pilot PN sequence
lengths. The feasibility of such simple channel estimation schemes 
that exploit the fade invariance in the delay-Doppler domain and the
robust detection performance even at high Dopplers (a feature that 
current multicarrier modulation schemes such as OFDM do not offer)
suggest that OTFS is a promising next generation modulation scheme 
suited for 5G and future wireless systems.

\end{document}